\pdfoutput=1
\documentclass[hidelinks,12pt,authoryear]{article}
\usepackage{float}
\floatstyle{plaintop}
\restylefloat{table}
\usepackage{physics}
\usepackage[caption = false]{subfig}
\usepackage[english]{babel}
\usepackage{textcomp}
  \usepackage{xltabular}
  \usepackage{threeparttable}
  \usepackage{threeparttablex}
\usepackage{amssymb}
\usepackage{longtable}
\usepackage{comment}
\usepackage{multirow}
\usepackage{amsmath, amssymb}
\usepackage{amsthm}
\usepackage{amsbsy}
\usepackage{hyperref}
\usepackage[font=footnotesize]{caption}
\usepackage{eurosym}
\usepackage[doublespacing]{setspace} 
\usepackage{inputenc}
\usepackage{geometry}
\usepackage{setspace}
\usepackage{multirow}
\usepackage{xy}
\usepackage{float}
\usepackage{eurosym}
\usepackage{natbib}
\usepackage{array,ragged2e}
\usepackage{threeparttable}
\usepackage{caption}
\usepackage{bbold}
\usepackage{amsmath, amssymb}
\usepackage{amsthm}
\usepackage{mathtools}
\usepackage[mathscr]{eucal}
\usepackage{amsbsy}
\usepackage[pdftex]{color,graphicx}
\usepackage{graphicx}
\usepackage{inputenc}
\usepackage{geometry}
\usepackage{booktabs}
\usepackage{hyperref}
\usepackage{nameref}
\restylefloat{table}
\usepackage{tabularx}
\usepackage{array}
\usepackage{tabulary}
\usepackage{pdflscape}

\newcommand{\R}{\mathbb{R}}
\newcommand{\e}[1]{{\mathbb E}\left[ #1 \right]}

\newcommand{\ou}{%
  \mathrel{%
    \vcenter{\offinterlineskip
      \ialign{##\cr$<$\cr\noalign{\kern-1.5pt}$>$\cr}%
    }%
  }%
}

\makeatletter
\newcommand\footnoteref[1]{\protected@xdef\@thefnmark{\ref{#1}}\@footnotemark}
\makeatother

\newgeometry{
 left=3.175cm, 
  right=3.175cm,
  top=3.175cm,
  bottom=3.175cm,
  bindingoffset=5mm
}

\usepackage[toc,page]{appendix}
\usepackage[bottom]{footmisc}

\begin{document}


	\title{\textsc{ \textbf{The Cost of Influence: \\ How Gifts to Physicians Shape Prescriptions and Drug Costs}}
	}

	\onehalfspacing
	\author{\sc Melissa Newham\thanks{Corresponding author. ETH Zurich and KU Leuven (e-mail: mnewham@ethz.ch).} \\ \sc Marica Valente\thanks{University of Innsbruck, Innsbruck Digital Science Center (DiSC), Innsbruck Decision Sciences (IDS), and DIW Berlin (e-mail: marica.valente@uibk.ac.at). We are very grateful to Elliot Ash, Albert Banal-Estanol, Tomaso Duso,  Shan Huang, Paolo Pertile, Julien Sauvagnat, Fiona Scott Morton, Jo Seldeslachts, Otto Toivanen, Hannes Ullrich and Frank Verboven for helpful comments. This paper also benefited from comments in seminars and conferences at the EuHEA Seminar Series, DIW Berlin, the ETH/UZH Seminar in Economics \& Data Science, UZH Business Department Seminar, the Zurich Political Economy Seminar Series and SESS Annual Congress. Parts of this research originate from a chapter of Newham’s PhD dissertation. A previous version of this paper was titled ``Who Pays for Gifts to Physicians? Heterogeneous Effects of Industry Payments on Drug Costs''. The authors declare that they have no relevant or material financial interests that relate to the research described in this paper. 
 }}

	\date{April 2023} 
 

	\maketitle

	\vspace{-1cm}
	\begin{abstract}

		\noindent \footnotesize{\noindent 

This paper studies how gifts -- monetary or in-kind payments -- from drug firms to physicians in the US affect prescriptions and drug costs. We estimate heterogeneous treatment effects by combining physician-level data on antidiabetic prescriptions and payments with causal inference and machine learning methods. We find that payments cause physicians to prescribe more brand drugs, resulting in a cost increase of \$30 per dollar received. Responses differ widely across physicians, and are primarily explained by variation in patients’ out-of-pocket costs. A gift ban is estimated to decrease drug costs by 3-4\%. Taken together, these novel findings reveal how payments shape prescription choices and drive up costs.


		}
		
		\vspace{1em}
		\vspace{-0.5cm}
		
	\end{abstract}
	
	\newpage
	
	\onehalfspacing
	\section{Introduction}

In 2020, drug companies in the US provided physicians with monetary and in-kind payments worth \$2 billion, including meals, gifts, consulting fees, and travel expenses.\footnote{US payment data is publicly disclosed since mid-2013 and is available from openpaymentsdata.cms.gov. The figure reported refers to general payments. Similarly in Europe gifts are frequent, however there is no transparency regarding their value \citep{fabbri2018sunshine}.}  
There is concern that these transfers can influence prescribing decisions and lead to costly and unnecessary prescriptions \citep{KING2017}. 
To reduce costs and avoid conflicts of interest in health care, a growing number of states have implemented, or considered, ``gift bans''.\footnote{The state of Vermont introduced a strict gift ban in 2009. California passed a gift limit in 2019 \citep{calif}, and the city of Philadelphia considered a ban in 2019 \citep{phil}.} Previous studies have shown that physicians are more likely to prescribe drugs for which they receive payments \citep{datta2017effects, carey2021drug, grennan2021no, agha2021drug}, however, the mechanisms underlying responses to payments are poorly understood. As policy effects may vary depending, e.g., on payment value, patient and physician characteristics, understanding the key factors driving physicians' responses to payments is crucial for evaluating the efficacy of gift bans

In this paper, we examine physicians' \textit{heterogeneous} responses to payments in terms of prescription decisions and drug costs for antidiabetic drugs in the US. 
To investigate the channels through which payments affect physicians' prescribing, we estimate responses to payments  using machine learning methods for flexible heterogeneous treatment effect estimation. 
Our new findings highlight one particular characteristic of physicians that is strongly associated with large responses to payments: treating patients who receive subsidies that lower out-of-pocket drug costs. Taking heterogeneity in responses into account, we quantify the impact of a gift ban on drug costs for certain states \textit{with and without} a ban. We estimate that a ban would reduce drug costs to treat diabetes by 3-4\%. 


Our analysis focuses on diabetes as it is one of the most costly common health conditions, with treatment expenses making up 20\% of all prescription drugs in the US \citep{zhou}. Further, treatment is primarily drug-based creating large incentives for pharmaceutical companies to influence physicians' prescriptions. There is also a growing concern that a large share of diabetic patients are being overtreated leading to hypoglycemia and increased risk of death \citep{bongaerts2021inappropriate, reuters}. Therefore, controlling drug costs and preventing overtreatment of diabetes is crucial for public policy. 

To evaluate the impact of payments on prescriptions, we combine data from Open Payments, a federal database on the universe of payments to physicians, with data on antidiabetic prescriptions and a large array of physician level characteristics from Medicare Part D over 2014-2017. We add relevant socioeconomic and health data for explaining antidiabetic prescriptions, such as local obesity and diabetes rates. Controlling for disease-specific covariates allows us to more cleanly identify the causal effects of payments on prescriptions.






We use machine learning-inspired matching estimators called causal forests \citep{atheyGRF} to consistently estimate physician level causal effects of payments with high-dimensional data. This is a powerful approach to estimating heterogeneous effects of payments (or ``treatment"), which accounts for the challenges posed by heterogeneity and selection (as drug firms may target specific physicians). 
One key advantage of using causal forests is  
to relax the assumption of constant treatment effects across physicians and estimate treatment effects at the individual level, making it possible to identify the key drivers of effect heterogeneity. Moreover, this nonparametric technique allows for the flexible modeling of both the outcome and selection equations, including complex interactions between covariates. Compared to standard OLS, this approach avoids arbitrary modeling choices and performs a data-driven search for parameter heterogeneity, which can be particularly valuable in cases where the sources of effect heterogeneity are potentially many and ex-ante unknown.



To ensure credible estimation of causal effects in this setting, it is crucial to meet some assumptions, namely, sufficient covariate overlap, unconfoundedness, and no spillovers \citep{wagerathey2019application}. To fulfill these assumptions, we first predict the propensity scores of receiving payments and limit the sample to comparable physicians who did and did not receive payments. Then, we apply the causal forest algorithm, which uses random forests to divide the covariate space into neighborhoods of physicians who are mostly similar in characteristics that influence responses to payments. We estimate treatment effects using residualized regression on similar physicians \citep{nie2019}, which allow to flexibly control for many possible sources of selection and outcome determinants, such as physician and year fixed effects and measures capturing potential spillover effects. We compare our estimator to standard OLS with fixed effects and evaluate the role of unobserved confounders using Oster's method (\citeyear{oster2019}). As we show, our estimates are unlikely to be overturned by unobserved confounders.



We find that, on average, a \$10 payment causes an \textit{increase} in the prescribed amount of brand drugs by 0.8 claims. Relative to the counterfactual, the average payment leads to a 4.8\% increase in brand prescriptions. We then analyze causal effects at the physician level. About 80\% of the estimated causal effects on brand prescriptions are positive and statistically significant. Effects differ widely across physicians, and range between 2\% to 11\%. This variation can be primarily attributed to one key characteristic of physicians, namely, treating a higher share of patients who qualify for a low-income subsidy that reduces out-of-pocket drug costs. 
This novel finding indicates that higher insurance amplifies the effects of payments. To a minor extent, we also find evidence that physicians with on average sicker patients respond more to payments. In contrast, we find no relationship between physician responses and the total value of payments received.

Our analysis reveals that about 20\% of payments intended to promote \textit{brand} drugs have positive spillover effects on the prescription of \textit{generic} drugs. This increase can be ascribed to the standard therapeutic approach to diabetes that entails a \textit{combination} of drugs from different classes, and the fact that major brand-name drugs in our sample have no bioequivalent generic substitute.

Next, we examine counterfactual scenarios with/without a gift ban for the Northern New England states of Maine, New Hampshire, and Vermont. This tri-state area has comparable health statistics in terms of medical expenses, healthcare costs and utilization, hospital statistics, and patient populations. 
Vermont is the only state in the US with a strict ban in place since 2009. Since the ban was implemented prior to the availability of payment data, it is not possible to gauge the effect of payments using a natural experiment approach. To tackle the issue of a missing control group within Vermont, we propose a policy evaluation approach that uses causal forests to predict counterfactual outcomes at the physician level.


Our counterfactual analysis focuses on the \textit{financial} impact of a gift ban. We find that, on average, for every dollar received, payments generate a \$30 increase in total drug costs. The gift ban in Vermont has resulted in savings amounting to 3.2\% (\$154k) of drug cost for diabetes for the period 2014-17. In the neighboring states of Maine and New Hampshire, a gift ban has the potential to decrease total costs by 4\% (\$630k) and 3.5\% (\$190k) respectively. 
For each patient, we estimate yearly potential cost savings of on average \$92 in Maine and \$55 in New Hampshire. 

From a policy perspective, the cost impact of a gift ban needs to be traded off against the potential effects of gifts on the quality of treatment. Most existing research does not point towards improvements in the quality of care as a result of gifts \citep{alpert2019origins, garcia2020medical, carey2021drug, bergman2021lobbying, agha2021drug}. In light of this, and given the mounting evidence of diabetes overtreatment, our findings raise the concern that payments may heighten the risk of excess medication.

This study contributes to the growing literature on industry payments to physicians. Prior research estimates the \textit{average} effect of payments on drug prescriptions using different samples and correction methods for selection bias (\citealp{datta2017effects}, for antiviral drugs; \citealp{carey2021drug}, for drugs of mixed therapeutic fields; \citealp{agha2021drug}, for anticoaugulants; \citealp{grennan2021no}, for statins). Our findings reveal that the mean payment causes a 4.8\% rise in prescription volume, which falls in line with estimates reported in previous research (5\% in \citealp{agha2021drug}; 1.6\% in \citealp{carey2021drug}). This study then goes further by using individualized treatment effects to show that responses to payments vary substantially across physicians. 

Our analysis of heterogeneity leads to new findings on the mechanisms behind responses to payments that were previously not considered in the literature. Specifically, our findings suggest that physicians take patients' out-of-pocket drug costs into account when responding to payments. This result contributes to the literature linking insurance with higher drug expenditures \cite[see e.g.][]{duggan2010effect, ghosh2019effect} and complements experimental evidence showing that  when physicians are provided with an incentive to promote drug sales, insured patients face higher drug costs than uninsured patients \citep{lu2014insurance}. 

Additionally, we show that payments targeting brand drugs can lead to spillover effects on prescriptions of \textit{generic} drugs which are often prescribed alongside brand drugs to treat diabetes. So far, the literature has investigated whether paid physicians switch patients to generic drugs as quickly as unpaid physicians \citep{carey2021drug}. Our finding adds a new insight showing payments' spillover effects from brand to generic drugs in markets where they can be complements. Finally, this study is the first to conduct a cost analysis at the physician level, revealing significant cost savings from a gift ban.

Our study also adds to the literature on industry payments from a methodological perspective. Prior research, as cited above, typically uses binary or categorical variables to represent payments and estimate their average effect on prescribing behavior. For the first time, our study utilizes recent developments in machine learning to estimate individualized treatment effects as a function of the payment amount received (continuous treatment) and a wide range of physician attributes. In doing so, we aim to provide more accurate estimates by leveraging the variability in payments and effectively accounting for many potential sources of response heterogeneity.

More broadly, this paper aims to contribute to the causal inference literature by using machine learning methods to estimate heterogeneous causal effects when individuals differ both in treatment values and observable characteristics. While successful implementations of machine learning for heterogeneous treatment effects in quasi-experimental settings can be found in environmental \citep{deryugina2019mortality, valente2023policy} and labor economics \citep{Pinotti}, applications in health economics are scarce \citep{kreif2021estimating}.  




The remainder of the paper is structured as follows. Section 2 describes background and data. Section 3 outlines our empirical strategy. Section 4 and 5 present the  results on prescriptions and drug costs, respectively. Section 6 concludes.

\section{Background and Data}

This section introduces the two main datasets used in the empirical analysis; Open Payments and Medicare Part D. Thereafter we provide information on the antidiabetic drug market, and introduce our final sample and key variables. 

\subsection{Open Payments}

Payments to physicians typically arise because of a face-to-face marketing encounter, known as a detailing visit.\footnote{Payments can also occur without any face-to-face encounter if, for example, a physician did not meet with the sales representative but still accepted a gift.} We obtain comprehensive data on payments to physicians from the Open Payments website. Open Payments is a national disclosure program created by the Affordable Care Act and managed by the Centers for Medicare and Medicaid Services (CMS) with the aim to promote transparency and accountability in the healthcare sector. The Act requires manufacturers to fully disclose payments of \$10  or more made to physicians and teaching hospitals. The first full year in which payments were publicly disclosed is 2014.\footnote{The 2013 program year includes only data collected from the second half of the year.}


While US law forbids companies from explicitly paying physicians to prescribe drugs or medical devices, gifts linked to the promotion of drugs or fees for consulting or speaking events are largely unrestricted. Only a handful of states ban certain types of payments or limit the value of such transfers.\footnote{States with statutory gift bans payments prior to the Sunshine Act include Vermont (2009), Colorado (2007), Minnesota (1994) and Massachusetts (2009) \citep{gorlach2013brightening}. Maine enacted a ban of certain types of payments at the end of our sample, in mid-2017, however clear rules only came into effect in June 2020 \citep{maine}. In our data we do not see a decline in the value or number of payments to physicians in Maine in 2017 vs. 2016 and 2015.} One of the most comprehensive statutory gift bans was implemented by Vermont in 2009. Vermont's gift ban (18 V.S.A \S 4631a) prohibits most gifts, including free meals, to physicians who regularly practice in Vermont. The law is actively enforced and violators have been made to pay penalties  in the past. In 2013, Novartis was reported to have paid \$36k because of six meals that violated the state's gift ban \citep{Hams2013}.


\subsection{Medicare Part D}
We assess the impact of payments on prescriptions made under Medicare Part D. We obtain data on prescriptions dispensed under the Medicare Part D Program from the Centers for Medicare and Medicaid Services (CMS).\footnote{The Medicare Part D Detailed Prescriber Public Use File (PUF) provides data on all prescriptions at the physician-drug-year level. A further database, the Medicare Part D Prescriber Summary PUF contains additional information at the physician-year level. In both datasets, physicians can be identified by their National Provider Identifier (NPI) and so the two datasets can be easily combined.} Medicare is the federal health insurance program in the US for people over the age of 65 and people with disabilities. Medicare Part D provides subsidized private insurance for outpatient prescription drugs for enrollees and represents about 30\% of US retail prescription drug expenditure \citep{kaiser2019}. 

Beneficiaries of Medicare Part D pay a share of drug costs themselves (“out-of-pocket”). This share depends on their plan. Beneficiaries elect either a Medicare Advantage Plan (MAPD) or a stand-alone drug plan (PDP). A MAPD provides more comprehensive cover and caps yearly out-of-pocket spending. Additionally, beneficiaries with sufficiently low income and assets receive a Low-Income Subsidy (LIS). In 2018, more than 12 million (30\%) people received low-income subsidies \citep{Cubanski2018}. Beneficiaries with a LIS benefit from zero or much lower out-of-pocket expenditures for premiums, deductibles and drug costs \citep{yala2014patterns}.\footnote{\cite{yala2014patterns} find that the average out-of-pocket costs for LIS beneficiaries are 74\% lower than of the out-of-pocket costs for non-LIS beneficiaries with gap coverage (\$148 vs. \$570).}


\subsection{Antidiabetics} \label{antidiab}

We focus on payments and prescriptions for drugs used to treat diabetes.
Diabetes is on the rise in the US, as well as globally, and costs to treat Medicare beneficiaries with diabetes have grown steadily overtime \citep{Cubanski2019}. Based on the full Medicare Part D dataset for the US, total Medicare expenditures for brand and generic treatments for diabetes amounted
to \$8.6 billion in 2013 and increased to \$17.6 billion in 2017. 




Diabetes has no cure and must be managed by life-long therapy. Treatment is primarily drug based, alongside comprehensive lifestyle modification. There are two main types of diabetes: type 1 and type 2. Type 1 patients are treated primarily with insulin. Type 2 diabetes, which accounts for 90-95\% of all cases, is treated in a more complex way. The first line medication is metformin which is used for as long as the body tolerates it, thereafter different drugs are added to the treatment regimen \citep{draznin20229}. Diabetes drugs can be grouped in several classes based on the drug mechanism of action (e.g. Insulins or DPP-4 inhibitors). SGLT-2 inhibitors is the newest drug class in our dataset, launched shortly before our sample begins (in 2013). 
Importantly, therapy often \textit{combines} drugs from different classes. The guidelines from the American Diabetes Association provides some suggestions of how this combination of drugs can proceed, however ultimately treatment is complex and depends on patient factors such as comorbidities, cardiovascular risk, weight, preferences for side-effects, tolerance and cost \citep{draznin20229}.  

There is no clear best drug class, owing to the different side effects associated with each class, patient factors, as well as the need for combination therapy. However, a growing number of medical studies have reported concerns of diabetes overtreatment \citep{reuters, maciejewski2018overtreatment,lipska2015potential, bongaerts2021inappropriate}. The consequences of overtreatment can be severe and include hypoglycemia, weight gain and cardiovascular diseases. Appendix A provides a more detailed discussion of the treatment of diabetes and provides an overview of the drugs in our sample and their respective drug class.



\subsection{Sample and summary statistics} 
Pharmaceutical companies make payments to physicians in connection with specific brand drugs. We aim to study heterogeneous responses to payments more generally and quantify the effect of banning all payments on drug costs. Consequently, we aggregate payments and prescriptions related to antidiabetics to the physician-year level. We aggregate by year so that payments can be matched with the publicly available prescription data from Medicare Part D  which is at the physician-drug-year level.
 We complement data on physicians, payments and prescriptions with socioeconomic and health information for the areas (5-digit zip code or county) in which physicians are located. This includes data on household income, education, and population demographics from the US Census Bureau's American Community Survey, and local obesity and diabetes rates from the Centers for Disease Control and Prevention.\footnote{Available at: https://gis.cdc.gov/grasp/diabetes/DiabetesAtlas.html}  A detailed description of the dataset construction can be found in Appendix \ref{data_app}. The list of all variables in the datasets, descriptions and sources is provided in Table \ref{var_def} in Appendix \ref{data_app}.
 


In order to obtain credible estimates of the causal effects of payments, we exclude physicians in Maine and New Hampshire that do not satisfy the overlap assumption when estimating propensities to receive payments.\footnote{All physicians must have close comparisons in the opposite treatment group (paid/unpaid) for unbiased estimates of the causal effects without relying on extrapolation \citep{ImbensRubin2015, ImbensRubin2015trimming}. We describe this assumption in detail in Section \ref{Sec:method}.} 
The resulting trimmed sample includes 1,720 physicians located in New Hampshire  (35\%), Maine (46\%) and Vermont (19\%). About 16\% of physicians in Maine and New Hampshire receive at least one payment for a total of \$27,239. Table \ref{tab:desc_payments} shows summary statistics for payments related to antidiabetic drugs. Variables are at the physician-year level unless otherwise stated. 



	\begin{table}[H] 
	\centering
	\footnotesize
		\caption{\footnotesize Summary statistics for payments related to antidiabetic drugs for paid physicians in Maine and New Hampshire over 2014-2017.}
			\label{tab:desc_payments}
			\begin{tabular}{lllllll}
	&\multicolumn{5}{c}{}   \\ \hline			
Obs. 414 & Median & Mean & Min. & Sd & Max. \\ 
  \hline
  No. payments per year & 3.00 & 7.49 & 1.00 & 10.50 & 64.00 \\  
  No. cash payments per year & 0.00 & 0.02 & 0.00 & 0.16 & 2.00 \\ 
  No. in-kind payments per year & 3.00 & 7.47 & 0.00 & 10.49 & 64.00 \\ 
  Value (\$)  payments per year & 30.09 & 65.79 & 0.83 & 82.09 & 521.66 \\ 
  Value (\$) cash payments per year & 0.00 & 0.63 & 0.00 & 4.85 & 62.50 \\ 
  Value (\$) in-kind  payments per year & 29.75 & 65.16 & 0.00 & 81.81 & 521.66 \\  \hline
		\end{tabular}
	\end{table}
\noindent	
The median paid physician receives three payments yearly for a total of \$30 with a range from about \$1 to \$500.\footnote{Each payment in Open Payments is linked to a specific drug, in cases where multiple drugs are listed the payment value is split equally amongst all listed drugs. This explains why the lowest payments in our sample are below the reporting threshold of 10\$ (see Appendix \ref{data_app}).} Physicians most often receive in-kind payments (98\%). Thus the typical transfer in our sample is a small in-kind payment, most frequently a meal. Looking at the time dimension, total payments are highest in 2014 (\$12,882), and of similar value in 2015 (\$3,736), 2016 (\$5,267), and 2017  (\$5,354). Moreover, there is no statistical difference in the distribution of payments across years or states (t-test, Wilcoxon rank sum, Kolmogorov-Smirnov: p-values$>$0.4). 

 \begin{table}[H] \centering
\footnotesize
\renewcommand*{\arraystretch}{1.1}\caption{\footnotesize
 Summary statistics for physicians receiving vs. not receiving payments, as well as physicians in Vermont (subject to a payment ban).}\resizebox{\textwidth}{!}{
 \label{tab:sum_stats}
\begin{tabular}{lll|ll|ll}
\hline
  & \multicolumn{2}{c}{No Payment} & \multicolumn{2}{c}{Received Payment} & \multicolumn{2}{c}{Vermont (ban)}  \\ 
 Variable & \multicolumn{1}{c}{Mean} & \multicolumn{1}{c|}{Sd} & \multicolumn{1}{c}{Mean} & \multicolumn{1}{c|}{Sd} & \multicolumn{1}{c}{Mean} & \multicolumn{1}{c}{Sd} \\ 
\hline
DRUG CLAIM COUNTS & &   &     & &     & \\
Brand drugs &  101.39 & 130.97 & 153 & 133.12 & 103.51 & 123.29 \\ 	
Generic drugs & 185.4 & 112.18 & 228.72 & 123.1 & 167.82 & 104.49 \\

 & &   & & &     & \\ 
TOTAL DRUG COSTS \$ & &   &     & &     & \\ 
Generic drugs & 3380.83 & 3671.74 & 3786.42 & 2972.93 & 2892.58 & 3001.29 \\
Brand drugs & 53081.85 & 79330.11 & 78250.01 & 69490.17 & 59786.83 & 92801.63 \\  

 & &   & & &     & \\ 
COVARIATES & &   &     & &     & \\ 
\textbf{Physician} & &     & & &     & \\ 
Family practitioner (0/1) & 0.62 & 0.49 & 0.69 & 0.46 & 0.61 & 0.49 \\ 
Male physician  (0/1) & 0.69 & 0.46 & 0.82 & 0.39 & 0.62 & 0.49 \\ 	
New practitioner  (0/1)  & 0.14 & 0.35 & 0.1 & 0.3 & 0.1 & 0.3 \\ 	
Antidiabetics claim share & 0.06 & 0.07 & 0.07 & 0.06 & 0.06 & 0.06 \\

 & &   & & &     & \\ 
\textbf{Physician's patients} & &     & & &     & \\
Share of beneficiaries $>65$ & 0.12 & 0.26 & 0.12 & 0.24 & 0.12 & 0.27 \\ 
Share of male beneficiaries & 0.43 & 0.1 & 0.45 & 0.08 & 0.43 & 0.12 \\ 
MAPD claim share & 0.2 & 0.15 & 0.2 & 0.13 & 0.09 & 0.08 \\ 
LIS claim share & 0.47 & 0.22 & 0.52 & 0.2 & 0.46 & 0.17 \\ 	
Average age of beneficiaries & 70.73 & 4.29 & 70.2 & 4.09 & 71.33 & 3.35 \\ 
Average risk score of beneficiaries & 1.2 & 0.26 & 1.17 & 0.2 & 1.13 & 0.22 \\ 	
Share of beneficiaries with insulin claims & 0.22 & 0.17 & 0.23 & 0.14 & 0.27 & 0.15  \\ 

 & &   & & &     & \\ 

\textbf{Physician's practice location}  & &     & & &     & \\
Population & 16462.21 & 11519.59 & 13235.54 & 9777.57 & 11205.96 & 7938.3 \\ 
Population per sq. mile & 920.21 & 1547.4 & 658.33 & 1124.07 & 643.63 & 1360.64 \\ 
Median household income \$ & 56600.39 & 19032.93 & 53798.09 & 16284.07 & 54799.75 & 12683.44  \\ 
No. diagnosed with diabetes & 13136.87 & 8168.4 & 12329.11 & 7508.85 & 4162.6 & 2114.85 \\   
Percent diagnosed with diabetes & 8.16 & 1.07 & 8.31 & 1.08 & 6.88 & 0.88 \\ 
No. obese  & 39000.49 & 24582.45 & 36071.12 & 22561.48 & 13216.86 & 7150.19 \\ 
Percent obese & 28.65 & 3.39 & 28.88 & 3.62 & 25.16 & 4.09 \\ 	
Population w/o high school degree $<24$  & 0.87 & 0.08 & 0.88 & 0.07 & 0.87 & 0.1 \\ 
Population w. college degree $<24$ & 0.55 & 0.14 & 0.52 & 0.16 & 0.52 & 0.18 \\ 
Population w/o high school degree 25-34 & 0.06 & 0.04 & 0.06 & 0.05 & 0.07 & 0.05 \\ 
Population w. college degree 25-34 & 0.36 & 0.17 & 0.32 & 0.17 & 0.39 & 0.18 \\ 
Population w/o high school degree 35-44 & 0.06 & 0.04 & 0.06 & 0.05 & 0.06 & 0.04 \\ 
Population w. college degree 25-34 & 0.36 & 0.17 & 0.31 & 0.17 & 0.42 & 0.16 \\ 
Population w/o high school degree 45-64  & 0.07 & 0.04 & 0.08 & 0.04 & 0.07 & 0.04 \\ 
Population w. college degree 45-64 & 0.31 & 0.14 & 0.27 & 0.13 & 0.37 & 0.12 \\ 
Population w/o high school degree $>65$ & 0.15 & 0.07 & 0.17 & 0.08 & 0.14 & 0.07 \\ 
Population w. college degree $>65$ & 0.27 & 0.12 & 0.23 & 0.11 & 0.33 & 0.12 \\ 
Share of White population & 0.93 & 0.05 & 0.94 & 0.05 & 0.94 & 0.04 \\ 	
Share of Black population & 0.02 & 0.02 & 0.01 & 0.02 & 0.01 & 0.01 \\ 	
Share of Asian population & 0.02 & 0.02 & 0.02 & 0.02 & 0.02 & 0.02 \\ 	
	

 & &   & & &     & \\ 

\textbf{Physician's network}  & &     & & &     & \\
No. of payments other physicians in zip code & 1.08 & 1.6 & 1.73 & 1.87 & 0 & 0 \\ 
No. of payments other physicians in county & 8.51 & 5.97 & 10.04 & 5.56 & 0 & 0\\ 	

No. of physicians in zip code & 11.92 & 10.23 & 8.49 & 7.17 & 7.17 & 6.06 \\ 
No. of physicians in county & 80.29 & 46.54 & 74.83 & 46.5 & 30.22 & 18.86 \\ 	 


 \hline
Observations (4712)    &  3279 &  3279  &  414 & 414 & 1019 & 1019 \\
\hline
\end{tabular}
}
\end{table}

Table \ref{tab:sum_stats} compares key attributes of untreated (No Payment) and treated (Received Payment) physicians in Maine and New Hampshire as well as physicians in Vermont for 2014-2017. It shows that paid physicians, on average, prescribe a higher volume of brand and generic drugs in comparison to unpaid physicians. Yet, large standard deviations relative to the mean indicate that the average effects of payments may mask important heterogeneities across physicians. 

Testing differences in means from Table \ref{tab:sum_stats} (t-tests) reveals that, on average, paid and unpaid physicians significantly differ along many dimensions. For instance, gender: about 82\% of paid physicians are male vs. 70\% for unpaid physicians. While this variable may be a good predictor of receiving a payment and prescribing brand drugs \citep[see, e.g.,][]{mendez2021gender}, it is still unclear whether it can drive differences in responses to payments. Table \ref{tab:sum_stats} also shows that average differences between paid and unpaid physicians are less pronounced for patient characteristics as well as socioeconomic and health data associated with practice location. Yet, as we describe in the next section, some of these variables may be important predictors of physicians’ responses to payments. 

Table \ref{tab:sum_stats} also shows that drug costs are higher for paid physicians. The magnitude of this gap raises several questions. After partialling out confounding factors, do reactions of physicians receiving high versus low payments differ? What role do patients' characteristics play in driving physicians' responses to payments? More broadly, how are prescribing patterns toward brand and generic drugs altered by payments?




\subsection{Relevant variables} \label{Sec:sel}


Our analysis controls for a large array of observables at the physician level that potentially affect prescriptions, payments, and responses to payments. Attributes can be grouped into four categories: physician characteristics, patient characteristics, characteristics of the practice location, and physician peer network.

\textit{Physician characteristics.} Prescribing patterns, payments and physicians’ responses to payments may differ depending on physician characteristics. For example, previous studies find that a physician's gender is an important predictor of differences in prescribing behavior \citep{tamblyn2003physician, zhang2019factors, mendez2021gender}. Prescribing patterns may also differ across general practitioners vs. specialists. For instance, payments may be more effective in persuading general practitioners to increase prescribing because they are less knowledgeable in a given area of specialty. Our analysis includes the following physician characteristics: physician gender, whether the physician is a new practitioner, a general practitioner or a specialist in diabetes, and the share of antidiabetic claims out of all claims which provides a further measure of degree of specialization.

\textit{Physician patient characteristics.} Patient characteristics such as age, gender, and general health status can affect the risk of diabetes and in turn influence payments and prescribing decisions. Drug firms may target physicians who treat a high volume of sick patients \citep{fugh2007following}. To investigate and control for differences in patient population we use information on Medicare beneficiaries, aggregated at the physician-year level. Variables include the average risk score (HCC)\footnote{The beneficiary average Hierarchical Condition Category (HCC) risk score is determined by CMS using demographic information and diagnoses on Medicare fee-for-service claims to measure each beneficiary's medical risk status, with higher scores going to beneficiaries with more (or more severe) health conditions or demographic risk factors. Thus, risk scores provide a proxy for patients' health status.}, 
the average age of beneficiaries, the share of beneficiaries over the age of 65, and the share of male beneficiaries. We adjust for the fact that physicians may see different proportions of patients with type 1 vs. type 2 diabetes by controlling for the share of beneficiaries with insulin claims as insulin is the only treatment option for type 1 diabetes. Previous studies indicate that physicians take patients' out-of-pocket expenditures for drugs and into consideration when prescribing drugs \citep[moral hazard, see][]{lundin2000moral} and their cost sensitivity  \citep{carrera2018physicians}.
Patient co-pays, and subsidies thereof, also affect decisions between brand and generic drugs \citep{dafny2017discounts}. In our analysis, patients’ out-of-pocket expenditures is proxied by information on patients’ insurance plans including the share of beneficiaries with a MAPD and a LIS. 



\textit{Physician practice location.} The geographic area in which a physician practices may influence prescriptions and payments. For example, if a physician practices in an area where the population has a higher risk of diabetes due to age, obesity and/or race, prescriptions and payments related to antidiabetic drugs may be higher.\footnote{Risk factors associated with type 2 diabetes, which accounts for 90-95\% of all diabetes cases, include being overweight, being 45 years or older, and being African American, Hispanic/Latino American, American Indian, or Alaska Native \citep{CDC2021}.} We control for disease-specific factors by including the number and percentage of adults diagnosed with diabetes, and the number and percentage of obese adults at the county level. 
To control for demographic and socioeconomic characteristics, we include information on population, population density, median household income and race in the 5-digit zip code area in which the physician's practice is located. \cite{bronnenberg2015pharmacists} find that how informed or expert consumers are affects their choice of brand in health care markets. To account for the potential influence of patients’ education we include several variables on educational attainment in the physician’s practice location (5-digit zip code level). 


\textit{Physician network characteristics.} Geographic proximity of physicians to other physicians has been shown to affect quality of care through increases in local competition \citep{ gravelle2019spatial, gravelle2016competition}. Further, drug firms may target physicians depending on the size of their network. To capture these possible confounding effects, we control for the network size of each physicians defined as the number of other physicians practicing in the same county and in the same 5-digit zip code. 

We also control for the number of payments to other physicians in the same network. Payments to peers may indirectly affect the prescribing behaviors of physicians belonging to the same peer network via referrals and other professional and social interactions \citep{agha2021drug}.\footnote{For full discussion of spillover effects, see Section \ref{Sec:method}.} Physicians are considered connected if they practice in the same 5-digit zip code or, more broadly, in the same county. 




\section{Empirical Strategy} \label{Sec:method}

Let ($y_{it},p_{it},x_{it}$) be the available data for physician $i=1,\dots,n$ at time $t=2014,\dots,2017$, where $x_{it} \in \R^d$ is a vector of $d$ exogenous covariates and fixed effects (year, physician, county), $y_i$ is the prescription outcome for either brand or generic drugs, and $p_{it} \in  [0;p_{max}]$ is the payment amount or ``treatment" variable in year $t$. Henceforth, we omit the true subscript $t$ for simplicity whenever possible. Note that payments range between zero (for all untreated physicians, namely, for all physicians not receiving any payment in a year) and the maximum payment set in treated physicians ($p_{max}$). 
We wish to estimate the target coefficient $\alpha_0$ in the model:
\begin{equation}
y_i = p_i\alpha_0 + x_i\beta_0 + \epsilon_i, 
\end{equation}
under the traditional assumptions of exogeneity of treatment assignment conditional on $x_i$, also known as unconfoundedness, and no spillovers (more precisely, the SUTVA or Stable Unit Treatment Value Assumption as in \citealp{Rosenbaum1983}).


We begin our analysis by estimating the average effect of payments, $\alpha_0$, using a standard OLS regression of brand drug prescription quantities on payments using our sample of treated (paid) and untreated (unpaid) physicians. The richness of our dataset allows to control for a large number of covariates and fixed effects. 


However, when the set of controls is very large and correlations between controls, treatment, and outcome variables are ubiquitous, the OLS estimator may return spurious estimates of the magnitude and significance of the treatment effect \citep{chern2018DML1, chern2018DML2}. To make the estimator robust to high-dimensional confounding, we proceed with a residualized regression approach. This approach is based on \cite{Robinson1988} and the Frisch-Waugh-Lovell theorem. In high-dimensional settings, this methodology has been generalized under the name of ``double/debiased machine learning" \citep{chern2018DML2, chern2018DML1}. Residualized regression recovers an estimate of $\alpha_0$ in three steps.





\begin{enumerate}
\item Consider the selection equation: $p_i = x_i'\pi_0^p + \gamma_i^p$ , 
and save the residuals $\gamma_i^p = p_i-\hat{p}_i$ resulting from partialling out the  effect of $x_i$ from $p_i$.
\item Consider the outcome equation:  $y_i = x_i'\pi_0^y + \gamma_i^y$ , 
and save the residuals $\gamma_i^y = y_i-\hat{y}_i$ resulting from partialling out the effect of $x_i$ from $y_i$.
\item Run the regression model:
 $\gamma_i^y =\alpha_0\gamma_i^p+\epsilon_i$ 
where $\alpha=\alpha_0$ solves the orthogonal population equation: $\e{(\gamma_i^y-\alpha\gamma_i^p)\gamma_i^p}=0.$ 

  \end{enumerate}
  \noindent



\noindent

In essence, residualization removes the correlation of covariates with payments and prescription outcomes, rendering the estimator robust to the parametric form in which covariates are included.\footnote{Residualization makes the estimator ``doubly robust”, i.e., as long as the estimator for either payments or outcomes is consistent, the resulting estimator for the treatment effect is consistent.} In comparison with OLS, double machine learning allows to flexibly control for a high-dimensional set of covariates by modelling (1) and (2) using machine learning algorithms optimized for prediction. 



To estimate heterogenous treatment effects we implement causal forests \citep{atheyGRF}. Similarly to other successful applications to quasi-experiments \citep[see, e.g.,][]{Pinotti}, we follow the implementation in \cite{atheyGRF}. Causal forests implement data-driven sample partitions that maximize treatment effect heterogeneity across sample splits. This creates neighborhoods of treated and untreated physicians who are mostly similar in characteristics that influence responses to payment. Treatment effects are estimated using residualized regression where observations are weighted by the frequency with which they end up in the same neighbourhood. As such, the estimator belongs to the class of adaptive (data-driven) $k$-nearest neighbor matching estimators. We refer to \cite{atheyGRF} for details. 

Credible identification of both average and heterogenous causal effects in this setting relies on the ``canonical'' assumptions of unconfoundedness, sufficient covariate overlap and no spillovers \citep{wagerathey2019application}. We proceed by discussing each assumption in turn. 

\textit{Unconfoundedness.} In order to fulfill unconfoundedness in the case of non-random payments, one needs to control for the sources of selection bias. The main concern in the literature is that drug companies target high volume prescribers \citep{carey2021drug, agha2021drug}. In this case, a cross-sectional comparison of paid vs. unpaid physicians would overstate the true effect of payments. In general, confounding factors that drive both brand prescriptions and the allocation of payments will give rise to selection concerns.






In this study we rely on the richness of our dataset, and it's panel structure, to control for selection. Thorough research and in-depth knowledge of our setting motivates the covariates included in $x_{i}$. We argue that we capture most measurable characteristics which drug companies themselves could observe when targeting physicians. Further, the application of nonparametric machine learning techniques allows to flexibly model both the outcome and the payment propensities, including complex interactions between covariates. 

To control for unobserved time-invariant physician characteristics, such as a preference for prescribing new or brand medications, we include physician fixed effects \citep[as implemented for identification in, e.g.,][]{datta2017effects, carey2021drug,agha2021drug}. Additionally, we control for common shocks to patient or physician preferences with year and county fixed effects. To control for changes in the targeting of payments and physicians' prescriptions over time, we include a large array relevant time-varying variables such as patients' average health risk and insurance coverage levels (described and motivated in detail in Section 2.5).\footnote{There is indeed large variation in, e.g., a physician's share of insured patients over time. Within physicians, the coefficient of variation of the share of patients with a MAPD insurance plan ranges between an average of 0.23 and a maximum of 1.25.} A potential omitted variable is payments received before the start of the dataset. We address this concern by adding the most recent lag of payments, when available, to the covariate set used to predict physicians' propensities to receive payments to capture possible path dependence in payments and reverse causality.  

After including all observed controls and fixed effects, addressing selection boils down to thinking about which unobserved time-variant confounders, not sufficiently proxied by the included covariates, remain and how large their impact could be. To gauge this, we implement the Oster (2019) method to place bounds on the degree of bias from unobservables. We show that bias-adjusted treatment effects are similar in magnitude to estimates obtained with residualized regression. 

\textit{Spillovers.} 
 Previous studies have shown that physicians may increase brand drug prescriptions in response to a peer physician receiving a payment due to the manner in which physicians may interact and influence each other \citep{agha2021drug}. Neglecting spillover effects from paid to unpaid physicians would thus \textit{understate} the direct impact of payments on prescribing. To address this concern, we control for the number of payments made to physician $i$'s peers in year $t$. One could define peer effects differently, e.g., based on the amount of shared patients, social interactions or closeness to particularly influential physicians. If more peer effects exist, our estimates would be a lower bound of the true effects of payments on prescriptions to the extent that peer effects enhance the impact of payments. 

\textit{Covariate overlap.} In order to obtain credible estimates of causal effects for each payment level, all paid physicians must have close comparisons
in the unpaid physician group 
\citep{ImbensRubin2015, ImbensRubin2015trimming}. Overlap in the estimated treatment propensities guarantees that a comparable unit can be found for each physician without relying on extrapolation. For a continuous treatment, the generalized propensity score (GPS) is the conditional probability of receiving a particular level of the treatment given the covariates \citep{Imbens2000}. When overlap in GPS is limited, one may consider trimming the sample and focusing on a subgroup with bounded propensity scores \citep{imbenslimitedoverlap}. This aims to improve the internal validity of our study because estimators for causal effects in the trimmed sample are more credible and accurate than estimators for causal effects in the original, full sample \citep{ImbensRubin2015trimming}. In our case, trimming the sample by removing observations with extreme values of the estimated propensity score (outside the 2.5$^{th}$ and 97.5$^{th}$ percentiles) ensures good overlap and allows for more robust inferences at the subsequent analysis stage. Details follow in Section \ref{sec:results}.

 \section{Results} \label{sec:results}



\normalsize
 \subsection{Average treatment effects of payments}

 Table \ref{tab:FE}  shows the average causal effect of payments on the amount of brand drug prescriptions. We present results including fixed effects to account for time-invariant heterogeneity across physicians and years.
 
 Rows (a)-(e) show the estimates of the average payment effect. Below the estimates, we present standard errors clustered at the physician level. Rows (a)-(b) are the standard OLS estimate with fixed effects and with/without controls described in Table \ref{tab:sum_stats}. Row (c) is the estimate from the residualized regression, as in \cite{chern2018DML1}. We estimate residualized drug prescriptions (outcome) and payments (treatment) flexibly using machine learning estimators that show high performance in terms of prediction accuracy (for the outcome) and propensity score overlap (for the treatment). These are the lasso estimator with fixed effects for the outcome and the regression forest estimator for payments. Row (d) shows the estimate of the average payment effect and its standard error  using the Augmented Inverse Propensity Weighing (AIPW) estimator, a doubly robust method that adjusts for covariates in the outcome model and the propensity score \citep{Robinson1994}. Row (e) is the estimate of the average payment effect computed using causal forests for heterogeneous treatment effects and taking the average of the estimated payment effects at the individual level for each physician.
 Row (f) is the lower bound of the average payment effect estimated following the adjustment strategy by \cite{oster2019} accounting for selection on unobservables. Row (g) is the  estimate of the average payment effect in c) multiplied by the mean payment and divided by the mean counterfactual outcome. It represents the percent increase in prescription volumes caused by payments. 

 

\begin{table}[H] 
\begin{center}
\caption{Average payment effect (APE) for 10\$ payment on brand drug claim counts.}
    \label{tab:FE}
\footnotesize
\begin{tabular}{lcc}
\hline\\[-1.8ex] 
  & Estimate \\ 
  & (Std. error)  \\ 
  \hline\\[-1.8ex] 
(a) APE OLS Fixed Effects - Uncontrolled coefficient  & 1.90 \\ 
   & (0.51)  \\ 
(b) APE OLS Fixed Effects - Controlled (full controls in Table  \ref{tab:sum_stats}) & 0.95  \\ 
   & (0.40) \\ 
(c)  APE Residualized regression (doubly robust) \citep{chern2018DML1}  & 0.78  \\ 
   & (0.22)  \\ 
(d) APE AIPW (doubly robust) \citep{atheyGRF} & 0.77  \\ 
   & (0.18)  \\ 
   (e) APE Causal forest \citep{atheyGRF} &0.79  \\ & (0.20) \\
(f) APE Bounded coefficient \citep{oster2019} & 0.80  \\

  \hline\\[-1.8ex]  
  (g) APE \% of counterfactual & 4.8\%  \\
     & (1.34)  \\
   
   \hline\\[-1.8ex] 
   No of obs treatment & 414  \\
   No of obs. control & 3279  \\
      \hline\\[-1.8ex]
      \end{tabular}
      \end{center}
      \vspace{-0.3cm}
 \footnotesize \textit{Notes:} All models include fixed effects (year, physician, county). P-values$<$0.01 for all coefficients. Regression standard errors are clustered by physician. (a) is the baseline estimate without controls (R$^2$ is 0.01).  (b) augments the baseline specification with the full set of controls (R$^2$ is 0.44). (c) estimates the effect using residualized regression where outcomes and payments are flexibly estimated via machine learning (lasso with fixed effects for outcomes and regression forests for payments). (d) is the estimate using the \textit{grf} implementation of the AIPW estimator (see \citealp{atheyGRF} section 4 for details). (e) shows the average of
the estimated payment effects for each physician using causal forests \citep{atheyGRF}. (f) reports the adjustment strategy by Oster (\citeyear{oster2019}) accounting for selection on unobservables.  We impose the most conservative estimate for $R_{max}^2$ (1.3$^*$R$^2$). (g) is the estimate in c) multiplied by the mean payment and divided by the mean counterfactual outcome. 
\end{table}
\noindent


We first run a simple OLS regression of payments on brand drug claims without controls (a). The point estimate of the effect of a \$10 payment is 1.9 with a standard error of 0.5. The effect decreases to 0.95 with a standard error of 0.40 when we control for our large set of covariates (b). The R$^2$ increases from 0.01 to 0.44. Hence, the uncontrolled coefficient displays upward bias, indicating that drug companies target on observable characteristics that are associated with higher brand prescription volumes. 

\begin{figure}[H]
    \centering
        \caption{Common support condition for physicians receiving (treated) and not receiving (untreated) payment.}    \includegraphics[width=16cm,height=5cm,keepaspectratio]{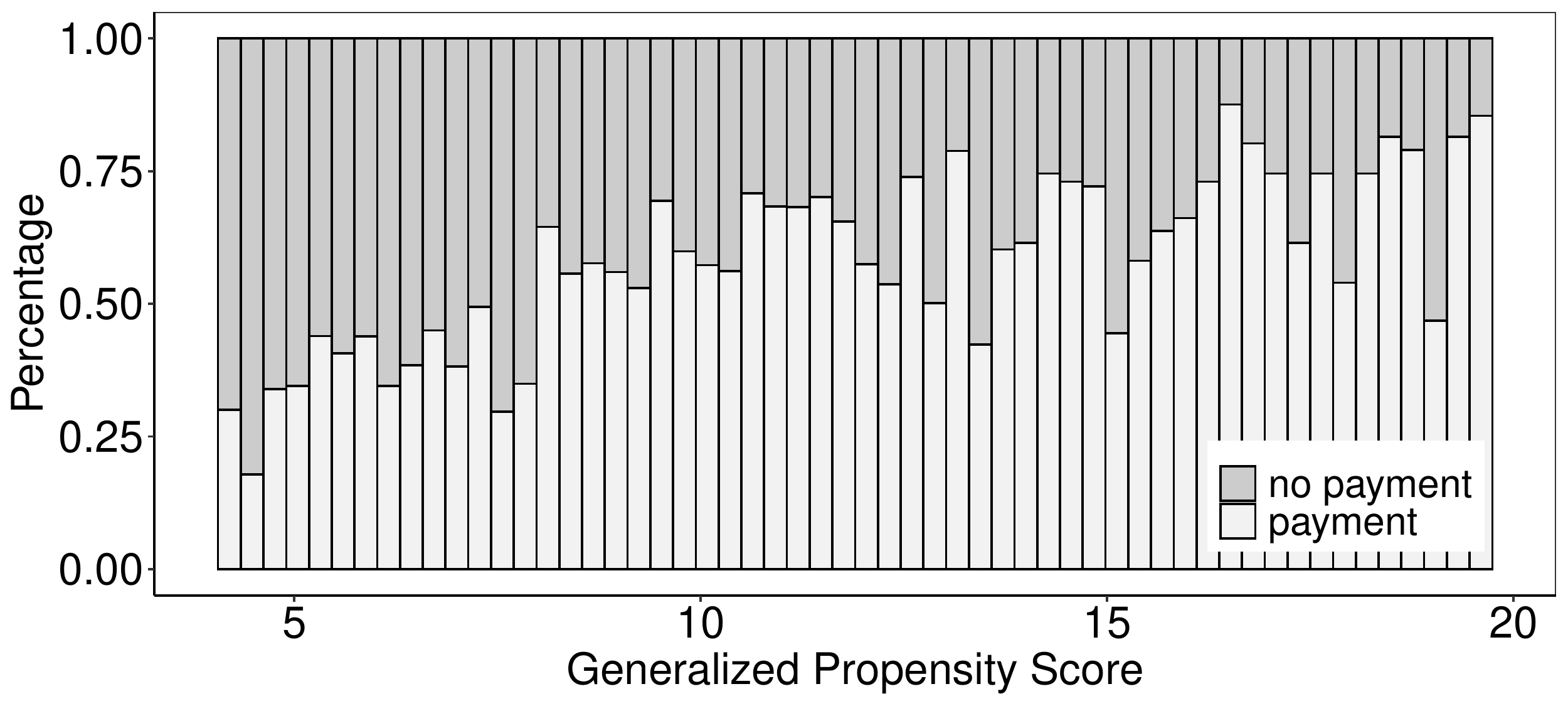}
    \label{fig:GPSoverlap}
\end{figure}

Next, we apply the residualized regression approach (c). We achieve good overlap in predicted propensity scores  across treatment groups as shown in Figure \ref{fig:GPSoverlap} and overlap statistics below. In addition, we achieve high accuracy in predicting brand drug prescriptions, as indicated by a correlation between predicted and observed outcomes of 98\% and as shown in Figure \ref{fig:y_vs_yhat} in Appendix. Estimation returns an average payment effect of 0.78. This indicates that payments significantly increase brand drug prescriptions -- if payments increase by \$10 relative to a trend, then the predicted brand drug claims go up by 0.78. This effect is the same when considering the {average treatment effect} and the average treatment effect \textit{on the treated}, which indicates that the estimation accounts well for the effect of selection. Next, we apply the AIPW estimator (d) and causal forests (e). These estimators return a similar result to (c). 


Finally, we compute the percent increase in prescription volumes caused by payments relative to the baseline mean counterfactual outcome (g). We find that the average annual payment is associated with a 4.8\% increase in prescription volume. This can be compared with an increase of 5\% as found by  \citet{agha2021drug} and the 1.6\% increase estimated by \citet{carey2021drug}.

\textit{Checking assumptions.} The estimates (c)-(e) rely on the assumptions of propensity score overlap (common support) and unconfoundedness. While the first can be tested empirically, the second cannot. One way to formally assess the role of unobserved confounders is to implement the method of Oster (\citeyear{oster2019}). This method takes into account movements in the R$^2$ across specifications with and without controls. We apply the suggested bounding exercise to the OLS point estimate of model (a) and (b). Table \ref{tab:FE} shows that the bounded coefficient (f) stays positive and close to the estimates (c)-(e) when assuming that selection on unobservables may be half the size of selection on observables in explaining the effects.

Moreover, we examine what amount of selection on unobservables in the opposite direction would be necessary to explain away the causal effect of payments on prescribing \citep[see][for implementation details]{oster2019}. This type of selection would require unaccounted-for unobservables that are positively correlated with payment and prescriptions. We find that omitted factors would need to be three times more strongly correlated with payments than all the controls accounted for in order to explain away the estimated effect of payments on brand drug prescriptions. This estimate implies an implausibly large degree of unaccounted-for selection, given the high-dimensional set of possible confounders we flexibly control for in (c)-(e). 
\normalsize

Concerning the assumption of common support, Figure \ref{fig:GPSoverlap} shows the distribution of predicted propensity
scores. The supports of the propensity scores across the two groups are similar after
trimming tail observations.\footnote{We trim observations with values outside the 2.5$^{th}$ and 97.5$^{th}$ percentiles of the distribution.} The resulting sample achieves good overlap statistics (summarized below), which indicates that we can estimate causal effects credibly for most
units without relying on extrapolation \citep{ImbensRubin2015, ImbensRubin2015trimming}.

   \vspace{-0.3cm}
   \noindent
 \footnotesize \textit{Notes:} Histograms of the estimates of the (generalized) propensity to be treated i.e. receive a given payment level. Estimates for paid physicians are in white, and for unpaid physicians in grey. Generalized propensity scores are estimated using regression forests of (continuous) payments on the full set of covariates including lagged payment amounts. \\

\normalsize
Four overlap statistics  give good summary
measures of covariate balance \citep{ImbensRubin2015}: the standardized difference in averages, the two sets of coverage frequencies,\footnote{I.e., the proportion of units with values within the 2.5$^{th}$ and 97.5$^{th}$ percentiles of the propensity score distribution.} and the number of units in each treatment group with a close comparison unit in the other treatment group. Assessing balance
 by inspecting these four measures reassures us about the presence of overlap. In particular, we find a small normalized difference in means (less than one standard deviation, 0.5), and high coverage proportions for the treated (93\%) and control group (90\%). Moreover, for all units we are able to find
close counterparts in the other treatment group, namely, we find at least one unit in the other treatment group with a less than 10\% difference in propensity scores.

\subsection{Heterogeneous estimates of payment effects}

\begin{figure}[H]
    \centering
    \includegraphics[scale=0.4]{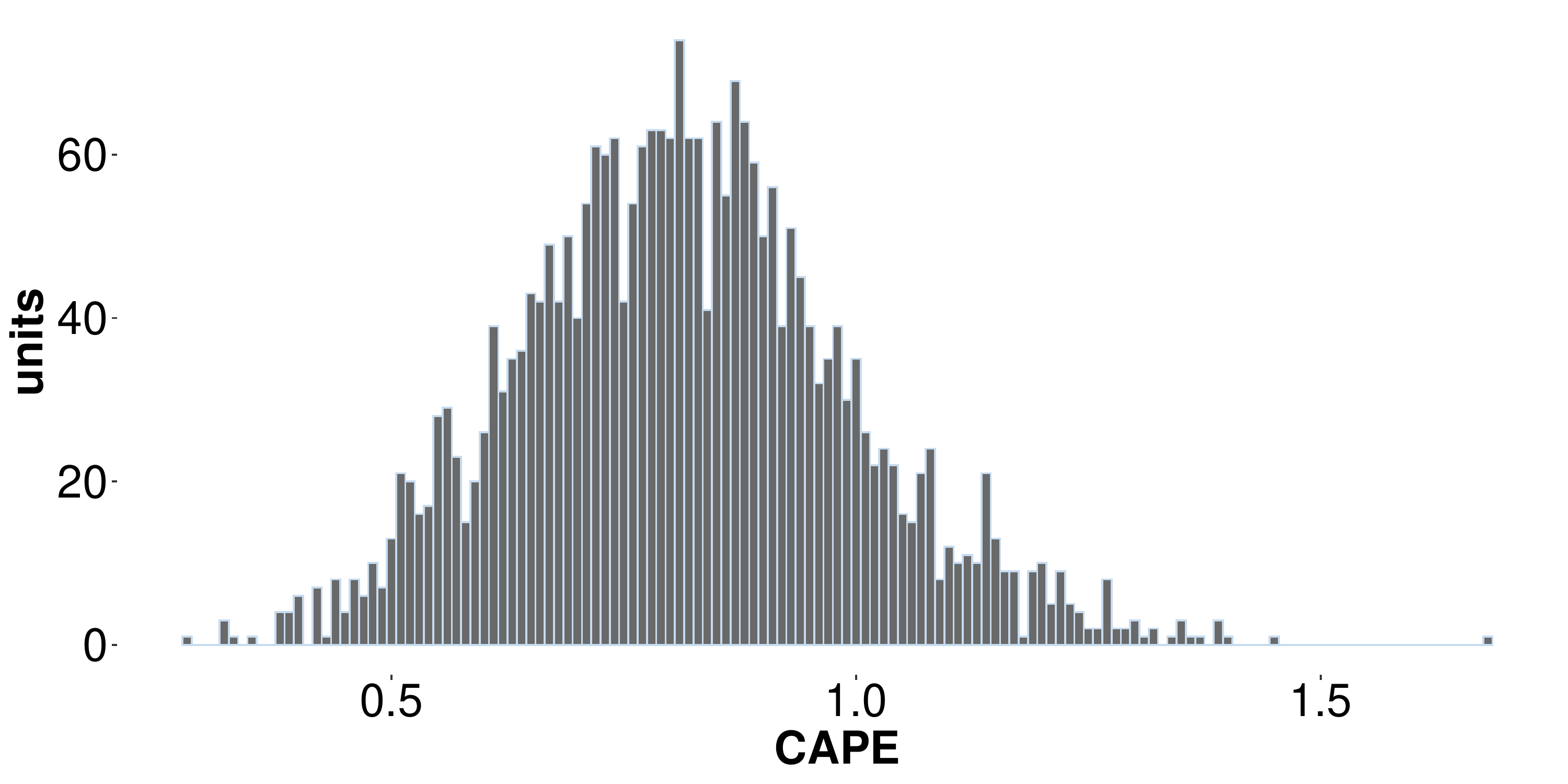}
    \caption{Physician level estimates of conditional average payment effects (CAPE) on brand drug prescriptions over 2014-2017. CAPE are measured as changes in brand drug claims for a \$10 increase in payment.}
    \label{fig:histcape}
\end{figure}

We now turn to the estimation and analysis of heterogeneous treatment effects. We apply the causal forest algorithm to estimate causal effects of payments on prescriptions of brand drugs at the physician level (CAPE). Estimation accounts for sources of selection using the predicted propensity scores in Figure \ref{fig:GPSoverlap} and, thereby, compares prescriptions of paid physicians with unpaid physicians with similar patients and personal attributes.\footnote{Implementation details can be found in Appendix \ref{app:implementation}.} We find that most payments (80\% or about \$19,200  in total) have a positive and statistical significant effect on claims of brand drugs (p-values$<$0.05), indicating a pervasive effect of payments on prescribing behavior. As shown in Table \ref{tab:FE} (e), on average, a \$10  payment increases the quantity of brand drugs by 0.79 claims.  This estimate masks considerable differences in responses. Responses differ widely, ranging between a 0.3 (2\%) and a 1.7 (11\%) increase in drug claims. Figure \ref{fig:histcape} presents the distribution of the estimated CAPE for all physicians in the sample showing statistically significant responses (p-values$<$0.05).\footnote{
The hypothesis of effect homogeneity is strongly rejected (Levene's test: p-value$<$0.01). We focus our attention on causal effects over the whole time period because dynamics are not specific to any particular year. Statistical tests (Kolmogorov-Smirnov, Mann-Whitney U) reject the hypothesis that the CAPE distribution differs across years (p-values$>$0.1).  Figure \ref{fig:histcapeyears} in Appendix plots the CAPE distribution by year.} 


\noindent

A simple metric of the importance of each variable for explaining heterogeneous treatment effects (CAPE) relates to the share of data-driven sample splits over a given characteristic \citep{wagerathey2019application}. 
We find that two groups of variables are used the most by the causal forest algorithm: the share of patients with low out-of-pocket expenditures (thus, with LIS and MAPD) and the health status of patients (thus, patients' age and average risk score). Hence we further explore these dimensions, as well as key physician characteristics that have been shown to affect prescription decisions in general (e.g. gender, specialization) and payment value. 


Table \ref{tab:CAPEcohen} compares the average characteristics of individuals with high and low treatment effects (CAPE), and formally tests for the difference in means while taking into account multiple hypothesis testing \citep{list2019}. Cohen (\citeyear{cohen1988}) suggested that standardized differences of 0.2, 0.5, and 0.8 can be used to represent small, medium, and large effect sizes, respectively. Our results confirm the importance of the share of patients with low out-of-pocket expenditures. In particular, one characteristic of physicians is especially associated with large effects of payments: treating patients who benefit from low-income subsidies (LIS). Differences in the share of patients with LIS is especially large (0.8). To a smaller extent, also differences in the share of patients with MAPD are significant and above the critical value (0.44). 

Moreover, Table \ref{tab:CAPEcohen} shows that payments have larger effects on prescriptions in the subgroup of physicians treating patients with a poorer health status, as indicated by significant differences in the average risk score (0.47). Taken together, these results show that physicians take their patients' financial expenditures and health status into account when making prescription decisions in response to payments. 


To a smaller extent, we also find effect heterogeneity depending on the type of physician, namely, responses are significantly larger in the group of family practitioners (vs. specialists). One potential explanation for this is that family practitioners may be less knowledgeable about antidiabetics and therefore more influenced by payments. Differently, there is no large heterogeneity by physicians' gender or for new practitioners. Finally, there is no heterogeneity in responses per dollar received depending on the total amount of payment. 

\begin{table}[H]
\centering
\footnotesize
\caption{Average characteristics of physicians with high vs. low CAPE (Conditional Average Payment Effect) on brand drug claim counts.}
\label{tab:MHT}
\begin{tabular}{lcccc}
  \hline\\[-1.8ex] 
   & (1)  & (2)  & (3) & (4) \\ 
   \cline{2-5}\\ [-1.8ex] 
 & CAPE$^{high}$ & CAPE$^{low}$ & Std. Diff. & p-value \\ 
  \hline\\[-1.8ex] 

 &  &  &  &  \\ 
\textbf{Patients' cost sensitivity} &  &  &  &  \\ 
LIS claim share & 0.54 &  0.38 &  0.80 & $<$0.01 \\
MAPD claim share & 0.23 & 0.16 & 0.44 & $<$0.01 \\ 
 &  &  &  &  \\
\textbf{Patients' health status} &  &  &  &  \\ 
Average risk score of beneficiaries & 1.27 & 1.15 & 0.47 & $<$0.01 \\  
Average age of beneficiaries & 70.5 & 71 & 0.11 & 0.07 \\ 
Percent diagnosed with diabetes (county) & 8.16 & 8.12 & 0.04 & 1.00 \\
 &  &  &  &  \\
\textbf{Physicians' type} &  &  &  &  \\ 
  Family practitioner & 0.63 & 0.52 & 0.22 & $<$0.01 \\  
  New practitioner & 0.14 & 0.12 & 0.08 & 0.39 \\ 
  Male physician & 0.72 & 0.72 & 0.01 & 1.00 \\ 
 &  &  &  &  \\
\textbf{Payment value} &  &  &  &  \\ 
Total value (\$) payments in year & 7.61
& 7.20 & 0.01 & 1.00 \\ 
   \hline\\ [-1.8ex] 
\end{tabular}
\label{tab:CAPEcohen}
\end{table}
\vspace{-0.5cm}
\noindent
\footnotesize \textit{Notes.} CAPE on brand drug claim counts is estimated using causal forests. Columns 1 and 2 report average characteristics of physicians with CAPE above and below the third and first quartile, respectively. Column 3 shows the magnitude of heterogeneity across groups measured as the standardized mean difference. Values of 0.2, 0.5, and 0.8 represent small, medium, and large heterogeneity between groups \citep{cohen1988}. Column 4 reports p-values testing for differences across groups with Holm-Bonferroni correction for multiple hypothesis testing \citep{list2019}.

\normalsize
\vspace{2em} 

Table \ref{tab:MHT} provides CAPE estimates per group of interest. The increase in brand drug prescriptions in response to payments is significantly higher (0.84 vs. 0.66, p-value$<$0.01) for physicians with a relatively larger share of patients receiving a LIS ($>$0.6). Concerning patients' health status, the CAPE is significantly larger  for physicians treating on average sicker patients (0.84 vs. 0.75, p-value$=$0.01) and younger  ($<$68 years old) patients (0.82 vs. 0.72, p-value$<$0.01). Responses also differ between family (0.82) and specialist (0.77) physicians (p-value$=$0.09).

In sum, our results indicate that the effects of payments are especially large in the group of physicians treating higher shares of patients with low out-of-pocket expenditures and sicker patients. Physicians' gender, experience, and amount of payment received do not significantly impact heterogeneity in payment responses.

\vspace{1em} 
\begin{table}[H]
\centering
\footnotesize
\caption{Estimates of Conditional Average Payment Effects (CAPE) on brand drug claim counts by physicians' characteristics.}
\label{tab:MHT}
\begin{tabular}{llccc}
  \hline\\[-1.8ex] 

Statistic &  Group & Value & S.E. & p-value \\ 
  \hline\\ [-1.8ex] 
    
 &  &  &  &  \\ 
&\textbf{Patients' cost sensitivity}  &  &  &  \\ 
CAPE &  Low LIS claim share  & 0.66 & 0.04 & $<$0.01 \\ 
 CAPE &  High LIS claim share & 0.84 & 0.02 & $<$0.01 \\ 
  Diff. CAPE & Low and high LIS claim share & 0.18 & 0.05 & $<$0.01 \\ 
   \hline\\[-1.8ex] 
CAPE &   Low MAPD claim share& 0.82 & 0.02 & $<$0.01 \\ 
CAPE &   High MAPD claim share & 0.81 & 0.02 & $<$0.01 \\ 
  Diff. CAPE & Low and high MAPD claim share & -0.01 & 0.03 & 0.88 \\ 
   
      &  &  &  &  \\
&\textbf{Patients' health status} &    &  &  \\ 
CAPE &   Low risk score of beneficiaries & 0.75 & 0.02 & $<$0.01 \\ 
CAPE &   High risk score of beneficiaries & 0.84 & 0.02 & $<$0.01 \\ 
  Diff. CAPE& Low and high risk score & 0.09 & 0.03 & 0.01 \\ 
      \hline\\[-1.8ex] 
      CAPE &   Low average age of beneficiaries & 0.82 & 0.02 & $<$0.01 \\ 
CAPE &    High average age of beneficiaries & 0.72 & 0.02 & $<$0.01 \\ 
   Diff. CAPE & Low and high average age of beneficiaries & -0.10 & 0.03 & $<$0.01 \\ 
      \hline\\[-1.8ex] 
CAPE &  Low share diagnosed with diabetes (county) & 0.85 & 0.02 & $<$0.01 \\ 
CAPE &    High share diagnosed with diabetes (county) & 0.84 & 0.01 & $<$0.01 \\ 
   Diff. CAPE& Low and high share diagnosed with diabetes & -0.01 & 0.03 & 1.00 \\ 

 &  &  &  &  \\
&\textbf{Physicians' type} &  &   &  \\ 
CAPE &   Specialist physicians & 0.77 & 0.03 & $<$0.01 \\ 
CAPE &  Family physicians & 0.82 & 0.01 & $<$0.01 \\ 
  Diff. CAPE& Specialist and family physicians & 0.05 & 0.03 & 0.09 \\ 
     \hline\\[-1.8ex] 
     CAPE &   Female physician & 0.81 & 0.03 & $<$0.01 \\ 
CAPE &   Male physician & 0.80 & 0.01 & $<$0.01 \\ 
  Diff. CAPE & Female and male physicians & -0.01 & 0.03 & 0.77 \\ 
     \hline\\[-1.8ex] 
CAPE &  Existing practitioner & 0.80 & 0.01 & $<$0.01 \\ 
CAPE &  New practitioner & 0.85 & 0.05 & $<$0.01 \\ 
 Diff. CAPE& Existing and new practitioners & 0.05 & 0.05 & 0.32 \\ 
  &  &  &  &  \\
&\textbf{Payment value}  &  &  &  \\ 
CAPE &   Low value (\$) payment received in year & 0.79 & 0.02 & $<$0.01 \\ 
CAPE &   High value (\$) payment received in year & 0.78 & 0.03 & $<$0.01 \\ 
  Diff. CAPE& Low and high value value (\$) payment  & -0.01 & 0.03 & 0.67 \\ 
   \hline\\ [-1.8ex] 
\end{tabular}
\label{tab:CAPEhighlow}
\end{table}
\vspace{-0.5cm}
\noindent
\footnotesize \textit{Notes.} CAPE on brand drug claim counts is estimated using causal forests \citep{atheyGRF}. Diff. CAPE refers to the difference between the conditional average treatment effects for the two groups. All groups are defined based on 0/1 values (for binary variables) or values above and below the third and first quartile (for continuous variables). P-values are for the null-hypotheses that the value is zero.\\

\normalsize

\subsection{Effects on generics}

To further understand the effects of payments on prescriptions, we analyze the effects of payments on generics.  In our data, payments are exclusively related to the brand drug classes where bioequivalent generic alternatives are not available (i.e. DPP-4 inhibitors, GLP-1 agonists, SGLT-2 inhibitors and insulins). Thus, for the treatment of diabetes during our sample period, there are very limited possibilities for physicians to directly substitute brand medications for generic medications. Rather, brand drugs are often prescribed \textit{in combination} with the first-line treatment metformin, which is classed as a sensitizer, or with drugs in the sulfonylureas class \citep{draznin20229}. Drugs in both of these classes are available as cheap generics. No new generics are launched during the sample period.

We expect that the effect of payments on generic drugs will depend on the extent to which generic drugs are \textit{substitutable or complementary} to brand drugs. Hence we estimate the effects of payments on generic drugs, splitting by drug class. We distinguish between three classes of diabetes treatments available in generic form: sensitizers and sulfonylureas, which are often combined with brand-name drugs and which together account for 97\% of generic prescriptions, and other generic drugs (meglitinides and alpha-glucosidase inhibitors).  

Table \ref{tab:FEgen} shows the estimates of the average payment effects (APE) on generic claim counts by drug class. We find that payments for brand drugs significantly increase the prescriptions of generic drugs, especially for treatments typically prescribed together with the brand drugs in our sample, i.e., sulfonylureas and sensitizers. Overall, the effects on generic drugs are small relative to the effects on brand drugs (cf. 0.2 to 0.8 in Table \ref{tab:FE}). Causal forest estimates show that only about 16-19\% of payments for brand drugs have a statistically significantly effect on sulfonylureas and sensitizers, showing that positive spillover effects of payments related to brand drugs on generic treatments is not pervasive across physicians.  The effects on all the other generics are close to zero and statistically insignificant. In sum, we only find positive effects of payments on generic drug prescriptions for generics that are complementary with 
brand drugs, and these effects are much smaller than the direct effects of payments on brand drugs. 



\begin{table}[H] 
\begin{center}
\caption{Average payment effect (APE) for 10\$ payment on claim count by drug class accounting for year and physician fixed effects.}
    \label{tab:FEgen}
\footnotesize
\begin{tabular}{lcc}
\hline\\[-1.8ex] 
  & Estimate \\ 
  & (Std. error)  \\ 
  \hline\\[-1.8ex] 

\textbf{Generic drug class sulfonylureas} &  \\ 
(a)  APE  Residualized regression \citep{chern2018DML1} & 0.21 \\ 
   & (0.10)  \\ 
      (b) APE AIPW \citep{atheyGRF} & 0.17  \\ 
   & (0.08)  \\
(c) APE \% of counterfactual & 2.2\%  \\
    & (1.09)  \\


(d) Share of significant effects (CAPE) \%  & 19\%  \\
     &   \\
   \hline\\[-1.8ex] 
   
\textbf{Generic drug class sensitizers} &  \\ 
(a) APE Residualized regression \citep{chern2018DML1} & 0.24  \\ 
   & (0.12)  \\ 
   (b) APE AIPW \citep{atheyGRF} & 0.23  \\ 
   & (0.10)  \\ 
  (c) APE \% of counterfactual & 1.3\%  \\
    & (0.64)  \\
(d) Share of significant effects  (CAPE) \%  & 16\%  \\
     &   \\
      
 \hline\\[-1.8ex] 
      
   No of obs treatment & 414  \\
   No of obs. control & 3279  \\
      \hline\\[-1.8ex]
      \end{tabular}
      \end{center}
      \vspace{-0.3cm}
      \noindent
 \footnotesize \textit{Notes:} The estimates of the average payment effects (APE) of a 10\$ payment on generic drug claim counts by drug class (sulfonylureas and sensitizers). All models include fixed effects. (a) reports the estimate using the doubly robust residualized regression estimator as described for Table \ref{tab:FE}.  Regression standard errors are clustered by physician. (b) shows the estimate using the doubly robust AIPW estimator as described for Table \ref{tab:FE}. (c) is the estimate in a) multiplied by the mean payment and divided by the mean counterfactual outcome. It represents the percent increase in prescription volumes caused by payments. (d) is the share of statistically significant effects (CAPE) estimated at the physician level using causal forests. 
\end{table}
\noindent

\section{Effects of payments on drug costs}

\subsection{Drug cost increases due to payments}

To determine the increase in drug costs due to payments we compute the increase in brand prescriptions for each physician receiving payment $p$, multiply this by the average unit cost per claim for brand drugs $c$, and take the sum over each physician $i=1,...,n$ with characteristics $x_i=x$: 

\begin{equation*}
\hspace{-0.5cm}
\text{Change in drug costs due to payments (\$)} = \sum_{i=1}^n \Big[\widehat{CAPE}(x)^*p^*c
\Big]
\end{equation*}

We estimate that payments cause physicians to increase brand drug prescriptions by a total of 1,528 claims and \$820,094 over 2014-2017 in the states of Maine and New Hampshire (see Table \ref{sum_stat_cost}). Given that payments totaled \$27,239 in this time period, the impact of \$1 in payments received by physicians on drug costs is \$30. Compared to a counterfactual world without payments, the estimated increase corresponds to a 3.5\% (New Hampshire) and 4\% (Maine) total cost increase for society to treat diabetes.\footnote{We obtain this figure considering only statistically significant increases in brand claims due to payments.} In other words, banning payments has the potential to decrease total costs to treat diabetes by up to 4\%. For each patient, this amounts to an average cost increase per year of \$92 in Maine and \$55 in New Hampshire.


\begin{table}[H]
\centering
\caption{Summary statistics for the effects of payments on brand claims and drug costs (\$). Estimates at physician level aggregated for Maine (ME) and New Hampshire (NH). \label{sum_stat_cost}}
\footnotesize
\vspace{1em}
\begin{tabular}{lllll}
  \hline
Outcome & State & Mean & Sd & Total \\ 
  \hline

  Brand claims & ME & 5.28 & 6.42 & 1130.81 \\ 
   & NH & 3.82 & 5.03 & 397.62 \\ 
   \hline
   Drug costs \$ & ME & 2944.38 & 3788.75 & 630097.30 \\ 
   & NH & 1826.90 & 2540.75 & 189997.21 \\ 
     \hline
\end{tabular}
\end{table}

\begin{figure}[H]
    \centering
    \includegraphics[scale=0.5]{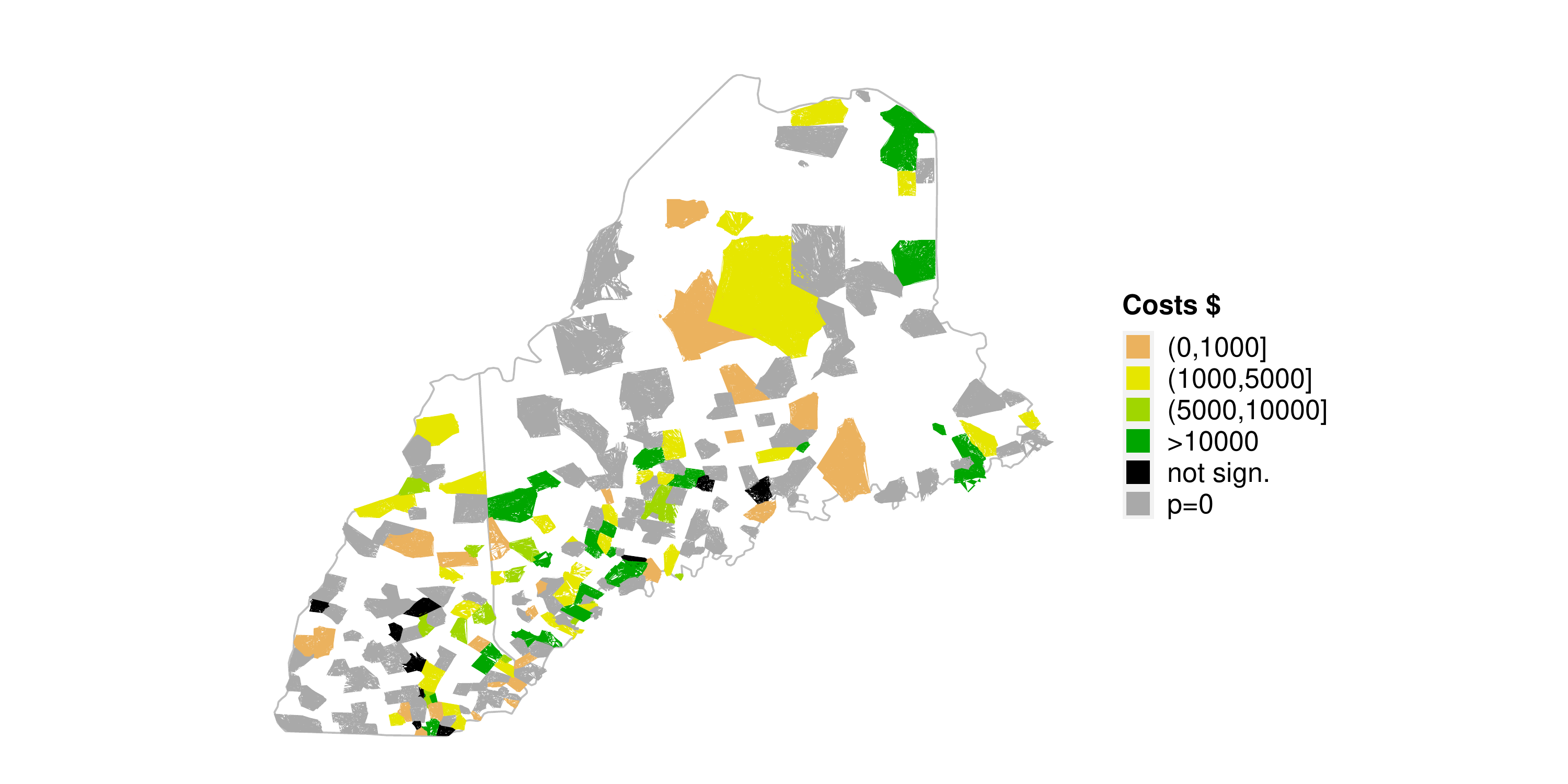}
    \caption{Drug costs (\$) increases due to payments. Physician level estimates aggregated by area (5-digit zip code).}
    \label{fig:sc}
\end{figure}

Figure  \ref{fig:sc} maps drug costs for physicians in Maine and New Hampshire, showing that estimates are heterogeneous within country and generally higher for physicians in Maine (dark green areas).

\subsection{The effects of a gift ban on drug cost}

We now quantify the cost savings in the period 2014-2017 due to the prevailing gift ban in Vermont. The first step entails answering the question: If all physicians in Vermont would be allowed to receive a payment, which amount would they receive? We answer this question by predicting propensities to receive payments for physicians in Vermont using the estimated function mapping covariates to predicted propensities in the other states. Then, we match payments of physicians in the other states with physicians in Vermont based on the nearest predicted propensities. Physicians in Vermont would have received a \$16 payment in median, for a total of about \$5,500. 

\begin{figure}[H]
    \centering
    \includegraphics[scale=0.5]{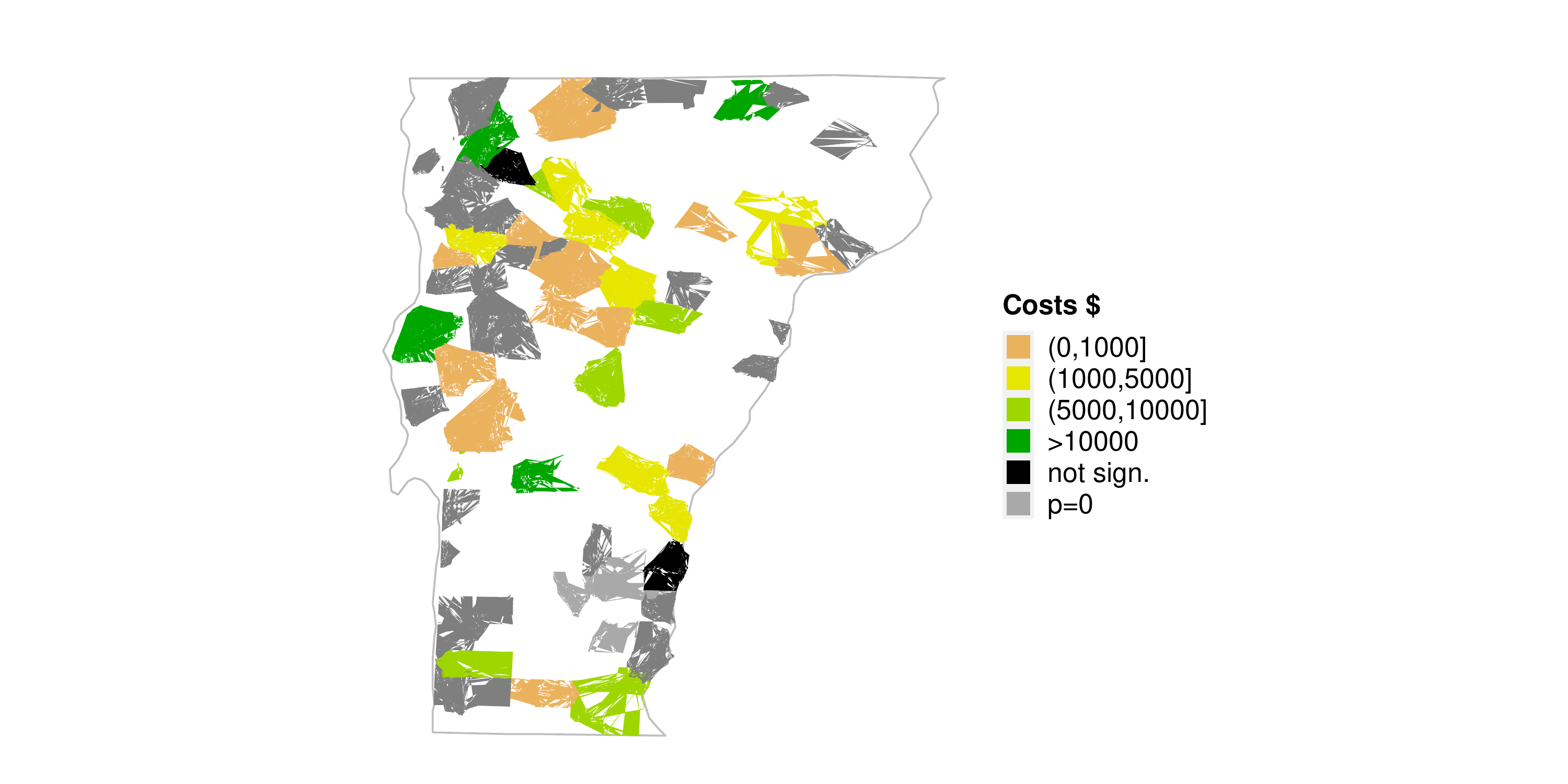}
    \caption{Predicted drug costs (\$) savings due to a ban on payments. Physician level estimates aggregated by area (5-digit zip code).}
    \label{fig:scVT}
\end{figure}

The second step entails answering the question: How would physicians in Vermont respond to such payments? We predict counterfactual responses to payments and corresponding costs for society using the estimated function via causal forests. We estimate that responses to payments are positive and statistically significant for about 80\% of physicians in Vermont. Overall, banning payments in Vermont has reduced brand drug prescriptions by 352 claims for a total cost saved of about \$154,039. Figure \ref{fig:scVT} maps the predicted drug cost savings in Vermont due to the gift ban. Although some physicians do not significantly respond to payments, aggregated drug cost savings at the local level (5-digit zip code) are positive overall. 

\noindent


The cost savings amount to 3.2\% of the actual costs for drug claims in Vermont. This figure is in line with the estimated share (3-4\%) of costs that New Hampshire and Maine would save with a ban on payments. This is reflects the fact that physicians in Vermont have similar characteristics to physicians in New Hampshire and Maine where payments are allowed. 

The implication of our analysis for policy is that banning payments has the potential to significantly lower drug costs. In addition to the financial impact of a gift ban, a further consideration for policy is whether and how payments affect the \textit{quality} of prescriptions. In fact, answering the question of whether paid and unpaid physicians prescribe drugs of a similar efficacy would require detailed health data at the patient level and medical expertise. While previous  studies generally find no evidence that payments lead to better health outcomes for patients \citep{alpert2019origins, carey2021drug, garcia2020medical, agha2021drug, bergman2021lobbying}, the health impacts of a gift ban may differ across therapeutic fields. For example, \cite{grennan2021no} find that in the case of statins where there is underprescribing, payments increase prescriptions to the benefit of patients. By contrast \cite{alpert2019origins} find that the introduction and marketing of OxyContin accounts for a substantial share of overdose deaths in the last two decades. In the case of diabetes, medical studies find that a substantial proportion of older adults with diabetes are potentially \textit{overtreated} and that the harms of intensive treatment likely exceed the benefits \citep{lipska2015potential, maciejewski2018overtreatment, bongaerts2021inappropriate}. The increase in both brand and generic prescriptions in response to payments may increase the risk of overtreatment. This provides some cause for concern that payments could harm patients with diabetes both financially and clinically. 

A gift ban may also cause equilibrium effects potentially leading to changes in drug pricing and direct to consumer advertising which in turn can impact on drug costs \citep{grennan2021no}. A full welfare analysis of a gift ban for antidiabetic drugs is outside of the scope of this paper, but is an important and challenging avenue for future research.

\section{Conclusion}


This paper examines how gifts from pharmaceutical firms to physicians in the US affect prescriptions and drug costs. Using recently developed methods at the intersection of causal inference and machine learning, we estimate heterogeneous treatment effects. We then identify the main factors driving responses to payments by analyzing variation in physician level responses. To quantify the potential savings from a gift ban, we develop a framework that uses causal forests to predict counterfactual outcomes at the physician level. 

Focusing on antidiabetic drugs prescribed under  Medicare  Part  D, we find that the average payment increases brand prescriptions by 4.8\%. Relative to the low
value of payments (a median of \$30 per year), the effects on drug costs to treat
diabetes are sizeable: On average, for every dollar received, payments generate a
\$30 increase in total drug costs.  Additionally, approximately 20\% of payments have positive spillover effects on generic drugs frequently prescribed alongside brand-name drugs.


Physicians differ widely in their responses to payments. Our analysis of heterogeneity in physician responses to payments yields three key insights: (i) The most important driver of physician responses is the insurance coverage of their patients. Specifically, physicians with a higher share of patients that benefit from a Medicare Part D low-income subsidy prescribe more brand drugs in response to payments. (ii) Physicians with on average sicker patients
respond more to payments. (iii) Payments of a higher value do not trigger stronger responses per dollar received. 

Finally, we find that a gift ban has the potential to reduce drug costs to treat diabetes by 3-4\%. The largest savings from a ban can be expected to come from lower prescriptions of brand drugs to patients whose drug costs are subsidized by Medicare Part D via low-income subsidies. Hence, in particular, the public healthcare system (and the taxpayer) will save on drugs costs. 


In conclusion, our research provides a set of facts which help us to understand how physicians respond to gifts from pharmaceutical companies. 
Our framework, which we apply to the market for antidiabetics, demonstrates that a gift ban could result in significant savings.  Further, our methodology can be directly applied to other states and therapeutic areas where data is available. 
Accurate information on the cost impact is a crucial part of cost-benefit trade off that policymakers need to make when deciding whether to ban gifts. The second part of the equation, understanding how gifts affect patients' health outcomes, is a question that we hope to tackle in future work.

\newpage

\normalsize

\appendix
\renewcommand\appendixpagename{Online Appendix}
\begin{appendices}
\section{Diabetes treatments} \label{app_anti_diab} 

\footnotesize{
Diabetes has no cure and must be managed by life-long therapy. There are two main types of diabetes: type 1 and type 2.  Type 1 patients are treated exclusively with insulin. Typically the treatment involves several insulins simultaneously. A patient can be treated with ``rapid'', ``long'' or ``intermediate'' insulin. An interview with a diabetologist revealed that physicians may change the type of insulin prescribed to a patient within these groups due to reasons such as side effects, patient intolerance, insurance reimbursement, and due to the introduction of new insulin that is generally perceived as a better product by the physician.  

Type 2 diabetes, which accounts for 90-95\% of all cases, is treated in a more complex way. The first line treatment is metformin (and comprehensive lifestyle modification) which is used for as long as the body tolerates it, thereafter different drugs are added to the treatment regimen \citep{draznin20229}. In patients with contraindications or intolerance to metformin, initial therapy should be based on patient factors \citep{draznin20229}. When monotherapy with metformin is no longer effective, physicians can choose between prescribing drugs from different drug classes for example DPP-4 inhibitors or SGLT-2 inhibitors. Type 2 patients with a severe condition are treated with insulin therapy.

Diabetes drugs can be grouped in several classes: sensitizers, insulins, GLP-1 receptor agonists, DPP-4 inhibitors, SGLT-2 inhibitors, alpha-glucosidase inhibitors, secretagogues and injectable amylin analogues. Drugs within the same drug class are more substitutable than drugs in different classes. However, therapy often combines several drugs from different classes. The guidelines for diabetes treatment from the American Diabetes Association provides some suggestions of how this step-wise addition of drugs can proceed, however ultimately treatment is complex and depends on patient factors such as comorbidities, cardiovascular risk, preferences for side-effects, tolerance and cost \citep{draznin20229}.  Below we provide an overview of the brand drugs and generics available in each drug class. The FDA approval year of each drug and generic entry year (if relevant) is provided in parentheses. The table provides a list of all diabetes treatments that are in the final dataset.

\noindent 

\begin{itemize}		
\item \textbf{Sensitizers}
	
	Biguanides/Metformin: Glucophage (1995), generic metformin hydrochloride (2003), Riomet (2003), Fortamet (2004), Glumetza (2005),  Actoplus Met (2005; generic 2011)
	

	
	TZDs (Thiazolidinediones): Avandia (1999), Actos (1999; generic 2012),  Avandamet (2002), Avandaryl (2005), Duetact (2006)
	
	\item \textbf{Insulins}
	
	Rapid and intermediate acting insulins: Humulin (1982), Novolin (1991), Humalog (1996), Novolog (2000), Apidra (2004), Afrezza (2014)
	
	Long acting insulins: Lantus (2000), Levemir (2005), Toujeo (2015), Tresiba (2015), Basaglar (2015), Xultophy (2016),  Soliqua (2016)
	
	\item \textbf{GLP-1 receptor agonists}
	
	
	Byetta (2005), Victoza (2010), Bydureon (2012), Tanzeum (2014), Trulicity (2014),  Adlyxin (2016)
	
	\item \textbf{DPP-4 inhibitors}

Januvia (2006), Janumet (2007),  Onglyza (2009), Kombiglyze XR (2010), Tradjenta (2011), Jentadueto (2012),  Nesina (2013), Oseni (2013), Kazano (2013) 
	
	\item \textbf{SGLT-2 inhibitors}

	
	Invokana (2013), Farxiga (2014), Invokamet (2014), Jardiance (2014), Xigduo (2014), Glyxambi (2015), Synjardy (2015)
	
	
	\item \textbf{Alpha-glucosidase inhibitors}

	Precose (1995; generic 2008), Glyset (1996)  
	
	\item \textbf{Secretagogues}
	
	Sulfonylureas: Glucotrol (1984, generic 1994), Glynase (1992), Glyburide Micronized (1992; generic 1997), Amaryl (1995; generic 2005), Glucovance (2000, generic 2004), Metaglip (2002; generic 2005)  
	
	
	Non-sulfonylurea secretagogues/Meglitinides: Prandin (1997, generic 2013), Starlix (2000, generic 2009), Prandimet (2008) 
	

	\item \textbf{Injectable amylin analogues}
	
	 Symlin (2005)
	
\end{itemize}
}

\begin{table}[htbp]\centering
		\footnotesize{
			\scalebox{0.85}{
				\begin{tabular}{lcc}			
					\hline			
					Firm and drug name	&	Drug class	\\ \hline
					PFIZER GLYSET	&	Alpha-glucosidase	\\
					GENERIC COMPANY PRECOSE	&	Alpha-glucosidase	\\
					ASTRAZENECA SYMLIN	&	Amylin agonists	\\
					MERCK JANUMET	&	DPP-4 inhibitors	\\
					BOEHRINGER INGELHEIM TRADJENTA	&	DPP-4 inhibitors	\\
					MERCK JANUVIA	&	DPP-4 inhibitors	\\
					TAKEDA KAZANO	&	DPP-4 inhibitors	\\
					BOEHRINGER INGELHEIM JENTADUETO	&	DPP-4 inhibitors	\\
					ASTRAZENECA ONGLYZA	&	DPP-4 inhibitors	\\
					ASTRAZENECA KOMBIGLYZE XR	&	DPP-4 inhibitors	\\
					TAKEDA OSENI	&	DPP-4 inhibitors	\\
					TAKEDA NESINA	&	DPP-4 inhibitors	\\
					GLAXOSMITHKLINE TANZEUM	&	GLP-1 agonists	\\
					NOVO NORDISK VICTOZA	&	GLP-1 agonists	\\
					ASTRAZENECA BYDUREON	&	GLP-1 agonists	\\
					ASTRAZENECA BYETTA	&	GLP-1 agonists	\\
					ELI LILLY TRULICITY	&	GLP-1 agonists	\\
					NOVO NORDISK TRESIBA	&	Insulins - Long	\\
					NOVO NORDISK LEVEMIR	&	Insulins - Long	\\
					ELI LILLY BASAGLAR	&	Insulins - Long	\\
					SANOFI TOUJEO SOLOSTAR	&	Insulins - Long	\\
					SANOFI LANTUS	&	Insulins - Long	\\
					ELI LILLY HUMULIN	&	Insulins - Rapid \& intermediate	\\
					NOVO NORDISK NOVOLOG	&	Insulins - Rapid \& intermediate	\\
					MANNKIND AFREZZA	&	Insulins - Rapid \& intermediate	\\
					ELI LILLY HUMALOG	&	Insulins - Rapid \& intermediate	\\
					NOVO NORDISK NOVOLIN	&	Insulins - Rapid \& intermediate	\\
					SANOFI APIDRA	&	Insulins - Rapid \& intermediate	\\
					BOEHRINGER INGELHEIM GLYXAMBI	&	SGLT-2 inhibitors	\\
					JANSSEN (SUB. PFIZER) INVOKAMET	&	SGLT-2 inhibitors	\\
					BOEHRINGER INGELHEIM SYNJARDY	&	SGLT-2 inhibitors	\\
					BOEHRINGER INGELHEIM JARDIANCE	&	SGLT-2 inhibitors	\\
					JANSSEN (SUB. PFIZER) INVOKANA	&	SGLT-2 inhibitors	\\
					ASTRAZENECA FARXIGA	&	SGLT-2 inhibitors	\\
					ASTRAZENECA XIGDUO	&	SGLT-2 inhibitors	\\
					NOVARTIS STARLIX	&	Secretagogues - Non-sulfonylureas	\\
					GEMINI LABORATORIES PRANDIN	&	Secretagogues - Non-sulfonylureas	\\
					GENERIC COMPANY STARLIX	&	Secretagogues - Non-sulfonylureas	\\
					GENERIC COMPANY PRANDIN	&	Secretagogues - Non-sulfonylureas	\\
					GENERIC COMPANY GLYBURIDE (MICRONIZED)	&	Secretagogues - Sulfonylureas	\\
					SANOFI AMARYL	&	Secretagogues - Sulfonylureas	\\
					GENERIC COMPANY GLYBURIDE BRAND	&	Secretagogues - Sulfonylureas	\\
					GENERIC COMPANY GLUCOTROL	&	Secretagogues - Sulfonylureas	\\
					BRISTOL MYERS SQUIBB GLUCOVANCE	&	Secretagogues - Sulfonylureas	\\
					PFIZER GLUCOTROL	&	Secretagogues - Sulfonylureas	\\
					GENERIC COMPANY METAGLIP	&	Secretagogues - Sulfonylureas	\\
					GENERIC COMPANY AMARYL	&	Secretagogues - Sulfonylureas	\\
					GENERIC COMPANY GLUCOVANCE	&	Secretagogues - Sulfonylureas	\\
					SANTARUS GLUMETZA	&	Sensitzers - Metformin	\\
					TAKEDA ACTOPLUS MET	&	Sensitzers - Metformin	\\
					ANDRX LABS FORTAMET	&	Sensitzers - Metformin	\\
					SUN PHARMACEUTICAL RIOMET	&	Sensitzers - Metformin	\\
					GENERIC COMPANY ACTOPLUS MET	&	Sensitzers - Metformin	\\
					GENERIC COMPANY METFORMIN BRAND	&	Sensitzers - Metformin	\\
					BRISTOL MYERS SQUIBB GLUCOPHAGE	&	Sensitzers - Metformin	\\
					TAKEDA ACTOS	&	Sensitzers - TZDs	\\
					GENERIC COMPANY ACTOS	&	Sensitzers - TZDs	\\
					\hline			
				\end{tabular}			
				
			}							
		}

		\caption{Diabetes treatments in the sample \label{all_treatments}
		}
\end{table}


\newpage

\section{Data Appendix} \label{data_app}

\subsection*{Dataset construction}

\footnotesize{The construction of the dataset follows four main steps. First, the set of all antidiabetic drugs is established. Next, we isolate prescription data from Medicare Part D on physicians that prescribe antidiabetic drugs in the states of Vermont, New Hampshire or Maine. In a third step, Part D prescriptions are matched with the payments data at the physician-drug-year level. Finally, the dataset is aggregated to the physician-year level. Below, each step is described in greater detail. 

The set of all approved treatments for diabetes (both brand and generic) is identified using the FDA Orange Book matched with Anatomical Therapeutic Chemical (ATC) Codes. We select drugs with the ATC ``A10 - Drugs used in diabetes''. The complete list of antidiabetic treatments is matched with the drug names in Part D and Open Payments using string-matching algorithms.

Information on the prescriptions of antidiabetic drugs, including the name and address of the prescribing physician, is extracted from the Medicare Part D database. This sample provides the universe of antidiabetic medications prescribed to patients enrolled in Medicare Part D. The sample is restricted to physicians located in Vermont, New Hampshire or Maine. 

Part D prescriptions are matched with the Open Payments data contained in the general payments file at the physician-drug-year level for the years 2014 to 2017. Each payment in Open Payments is linked to a specific drug, in cases where multiple drugs are listed the payment value is split equally amongst all listed drugs. This explains why the lowest payments in our sample are below the reporting threshold of 10\$. The dataset is aggregated to the physician-drug-year level such that payment values reflect the sum of all payments associated with a specific drug in a given year. The data is matched with Part D on the basis of drug name and physician name. Since there is no common physician ID that connects Part D and Open payments\footnote{The Part D data uses each provider's NPI number as its unique ID. The Open Payments system uses a randomly generated unique ID.}, the datasets are matched on the basis of full name and 9-digit zip code. This is complemented by a manual check in cases where physicians in Part D did not directly match to the Open Payments database.

Lastly, the dataset at the physician-drug-year is aggregated to the physician-year level. The final sample is restricted to physicians. Nurses and physician assistants are dropped due to the fact that they never receive payments from pharmaceutical companies. Observations with
extreme payments are dropped (above the 95$^{th}$ percentile). Observations with missing information for any variable are dropped. Certain information in Part D is redacted for cases where a drug is prescribed to 10 or fewer unique beneficiaries, thus physicians who prescribe antidiabetic medication to a total of 10 or fewer beneficiaries per year end up being dropped.\footnote{These physicians account for only 8\% of all claims in the data, thus the final dataset covers the majority of prescriptions by physicians for diabetes treatments.} Despite the ban, a small fraction (4\%) of physicians in Vermont received a positive payment at some point in the time span 2014 to 2017. These observations were excluded from the analysis.}

\newpage


\begin{landscape} 
\tiny
\setlength\LTleft{0pt}            
\setlength\LTright{0pt}           
\begin{longtable}{@{\extracolsep{\fill}}ll} 

 \endfirsthead
 \endhead
    \hline \multicolumn{2}{r}{{Continued on next page}} 
\endfoot
\endlastfoot
  \caption{List of variables, description and sources \label{var_def} } \\
  \label{table:des_appendix} \\[-2.8ex]
 \hline
Variable	&	Description (Source)	\\	\hline
PAYMENTS	&				\\
No. payments	&	Total number of payments for physician per year	(Open Payments)	\\
No. in-kind payments 	&	Total number of in-kind payments for physician per year		(Open Payments)	\\
No. cash payments 	&	Total number of cash payments for physician per year (Open Payments)	\\
Value (\$) payments 	&	Value (\$)  of  payments for physician per year (Open Payments)	\\
Value (\$) in-kind payments 	&	Value (\$) of in-kind payments for physician per year	(Open Payments)	\\
Value (\$) cash payments 	&	Value (\$) of cash payments  for physician per year (Open Payments)  \\
	&				\\
DRUG CLAIM COUNTS	&				\\
Generic drugs 	&	Number of Medicare Part D claims for generic antidiabetic drugs prescribed by physician per year (Part D Detailed Data) 	\\
Brand drugs 	&	Number of Medicare Part D claims for brand antidiabetic drugs prescribed by physician per year (Part D Detailed Data)	\\  
	&				\\
TOTAL DRUG COSTS 	&				\\
Generic drugs 	&	Total yearly drug cost for generic antidiabetic prescribed by physician. The drug cost includes the amount paid by the Part \\
& D plan, the beneficiary, government subsidies, and any other third-party payers. 		
(Part D Detailed Data) \\
Brand drugs 	&	Total yearly drug cost for brand antidiabetics prescribed by physician. The drug cost includes the amount paid by the Part \\ 
& D plan, the beneficiary, government subsidies, and any other third-party payers	
(Part D Detailed Data)			\\
	&				\\
COVARIATES	&					\\
Family practitioner (0/1)	&	Indicator taking the value 1 if speciality is recorded as Family Practice, Family Medicine and/or General Practice (vs. Specialist practitioner) \\
& (Part D Prescriber Summary, Open Payments)	\\
Male physician  (0/1)	&	Indicator taking the value 1 if physician gender is male 	(Part D Detailed Data)	\\
New practitioner  (0/1) 	&	Physicians with a provider enumeration year (date of NPI assignment) later than or equal to 2008	(NPPES NPI Registry)	\\
Antidiabetics claim share 	&	Share of claims for antidiabetic drugs out of all claims	(Part D Detailed Data)	\\
Share of beneficiaries $>65$ 	&	Share of beneficiaries age 65 and older with at least one claim for an antidiabetic drug 	(Part D Prescriber Summary)	\\
Share of male beneficiaries 	&	Share of male beneficiaries	(Part D Prescriber Summary)	\\
MAPD claim share 	&	Share of total claims attributable to beneficiaries covered by MAPD plans	(Part D Prescriber Summary)	 \\
LIS claim share 	&	Share of total claims attributable to beneficiaries with a Part D low-income subsidy	(Part D Prescriber Summary)	\\
Average age of beneficiaries	&	Average age of beneficiaries (Beneficiary age is calculated at the end of the calendar year or at the time of death)	(Part D \\
& Prescriber Summary)	\\
Average risk score of beneficiaries 	&	Average Hierarchical Condition Category (HCC) risk score of beneficiaries determined based on beneficiaries' diagnoses \\ & or demographic risk factors (Part D \& Prescriber Summary)	\\
Share of beneficiaries with insulin claims 	&	Share of claims for insulin out of all claims for antidiabetic drugs	(Part D Detailed Data)	\\
Population 	&	Total population in physician practice location (5-digit zip) (American Community Survey, ACS)	\\
Population per sq. mile	&	Total population divided by area in square miles for physician practice location (5-digit zip)	(ACS)	\\
Median household income 	&	Median household income in the past 12 months in dollars for physician practice location (5-digit zip) (ACS)	\\
No. diagnosed with diabetes 	&	Number of adults (20+) with diagnosed diabetes in physician practice location (county) \\
& (Centers for Disease Control and Prevention, CDC)	\\
Percent diagnosed with diabetes 	& Age-adjusted percentage of adults (20+) with diagnosed diabetes in physician practice location (county) (CDC) \\
No. obese 	&	Number of adults (20+) with obseity in physician practice location (county)	(CDC)	\\
Percent obese 	&  Age-adjusted percentage of adults (20+) with obesity in physician practice location (county)	(CDC)	\\
Population w/o high school degree $<$ 24 	&	Share of population below 24 years with less than high school degree in physician practice location (5-digit zip)  (ACS)\\
Population w. college degree $<$ 24  	&	Share of population below 24 years with some college or associate's degree in physician practice location (5-digit zip)  (ACS)	\\
Population w/o high school degree 25-34	&	Share of population 25 to 34 years with less than high school degree in physician practice location (5-digit zip) (ACS)	\\
Population w. college degree 25-34		&	Share of population 25 to 34 years with some college or associate's degree in physician practice location (5-digit zip)  (ACS)	\\
Population w/o high school degree 35-44	&Share of population 35 to 44 years with less than high school degree in physician practice location (5-digit zip) (ACS)		\\
Population w. college degree 35-44		& Share of population 35 to 44 years with some college or associate's degree in physician practice location (5-digit zip)  (ACS)	\\
Population w/o high school degree 46-64	&	Share of population 46 to 64 years with less than high school degree in physician practice location (5-digit zip)  (ACS)		\\
Population w. college degree 46-64	&	Share of population 46 to 64 years with some college or associate's degree in physician practice location (5-digit zip) (ACS)	\\
Population w/o high school degree $>$ 65	&	Share of population above 65 years with less than high school degree in physician practice location (5-digit zip) (ACS)		\\
Population w. college degree $>$ 65 	& Share of population above 65 years with some college or associate's degree in physician practice location (5-digit zip)  (ACS)	\\
Share of White population 	&	Share of White population in physician practice location (5-digit zip)	(ACS)	\\
Share of Black population 	&	Share of Black population in physician practice location (5-digit zip) 	(ACS)	\\
Share of Asian population 	&	Share of Asian population  in physician practice location (5-digit zip) 	(ACS)	\\
Share of Multi-race population	&	Share of Multi-race population  in physician practice location (5-digit zip) 	(ACS)	\\
No. of payments other physicians in zip code	&	Number of total payments to other physicians in sample in same 5-digit zip area and year	(Open Payments \& Part D Detailed Data)	\\
No. of payments other physicians in county 	&	Number of total payments to other physicians in sample in same county and year (Open Payments \& Part D Detailed Data)	\\
No. of physicians in zip code	&	Number of other physicians in sample in same 5-digit zip area	(Part D Detailed Data)	\\
No. of physicians in county 	&	Number of other physicians in sample in same county 	(Part D Detailed Data)	\\
\hline
\end{longtable}
\end{landscape}

\newpage

\newpage

\newpage
\section{Implementation details} \label{app:implementation}
We have explored from 500 to 20,000 trees in the RF, and treatment effect estimates become stable after 15,000 trees, thus, results are obtained using this value. All trees are grown with cross-validated values for the number of randomly subsampled covariates, minimum leaf size, and penalty for imbalanced splits, namely, splits in which the size of parent and child node are very different are penalized. In particular, each node is required to include a minimum number of both treated and control units, i.e., enough information about both factual and counterfactual to estimate the treatment effect reliably. For this reason, a penalization is imposed also to nodes including an unbalanced number of treated and control units. Following \citet{atheyGRF}, values for such parameters are obtained via cross-validation.\footnote{We use the software R-4.2.1 (2022) and the \textit{grf} package version 1.2.0 \citep{grfpack}.} 

\newpage

\newpage 
\section{Other results}\label{app:pred_Y}

\begin{figure}[H]
    \centering
    \includegraphics[scale=0.3]{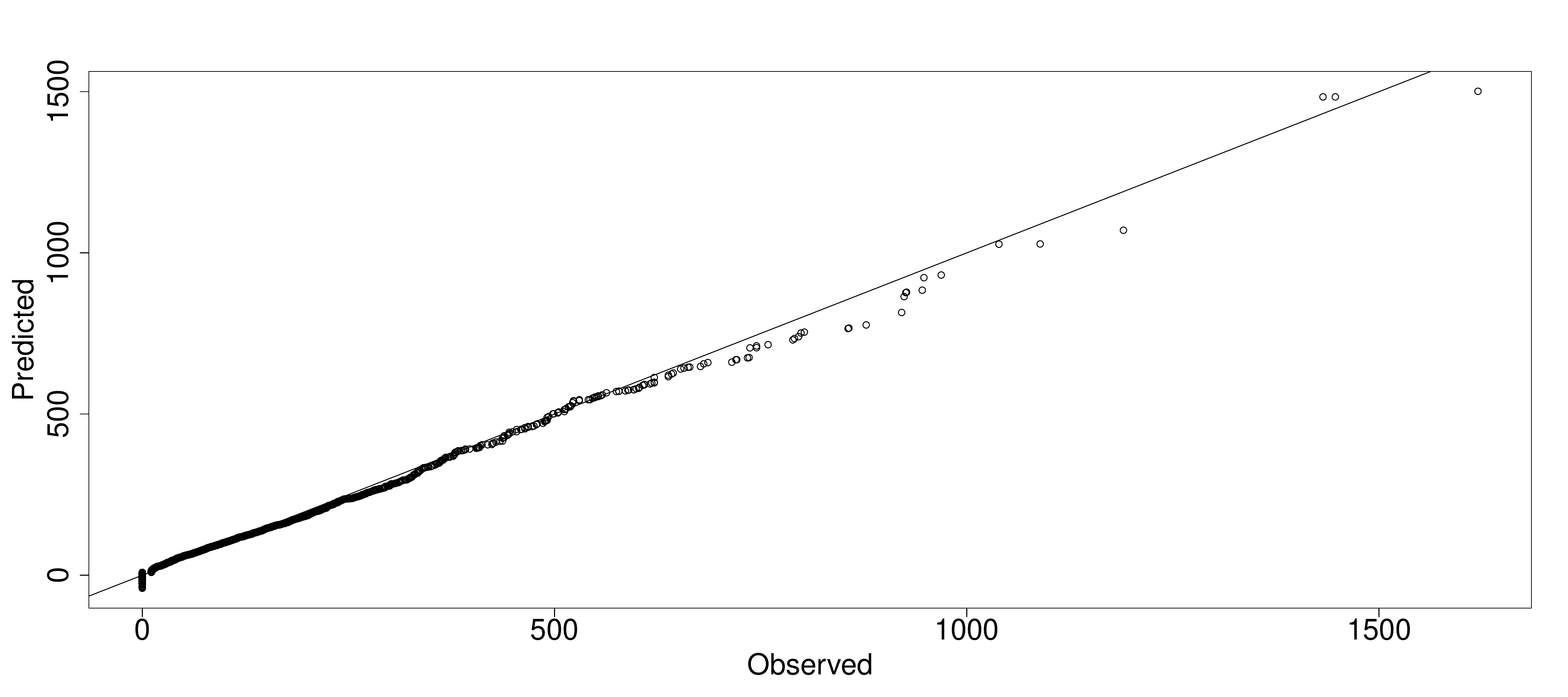}
    \caption{\footnotesize Outcome predictions applying the lasso estimator on the full set of controls and fixed effects to account for time-invariant variation across physicians and years.}
    \label{fig:y_vs_yhat}
\end{figure}
\noindent

\begin{figure}[H]
    \centering
    \includegraphics[scale=0.4]{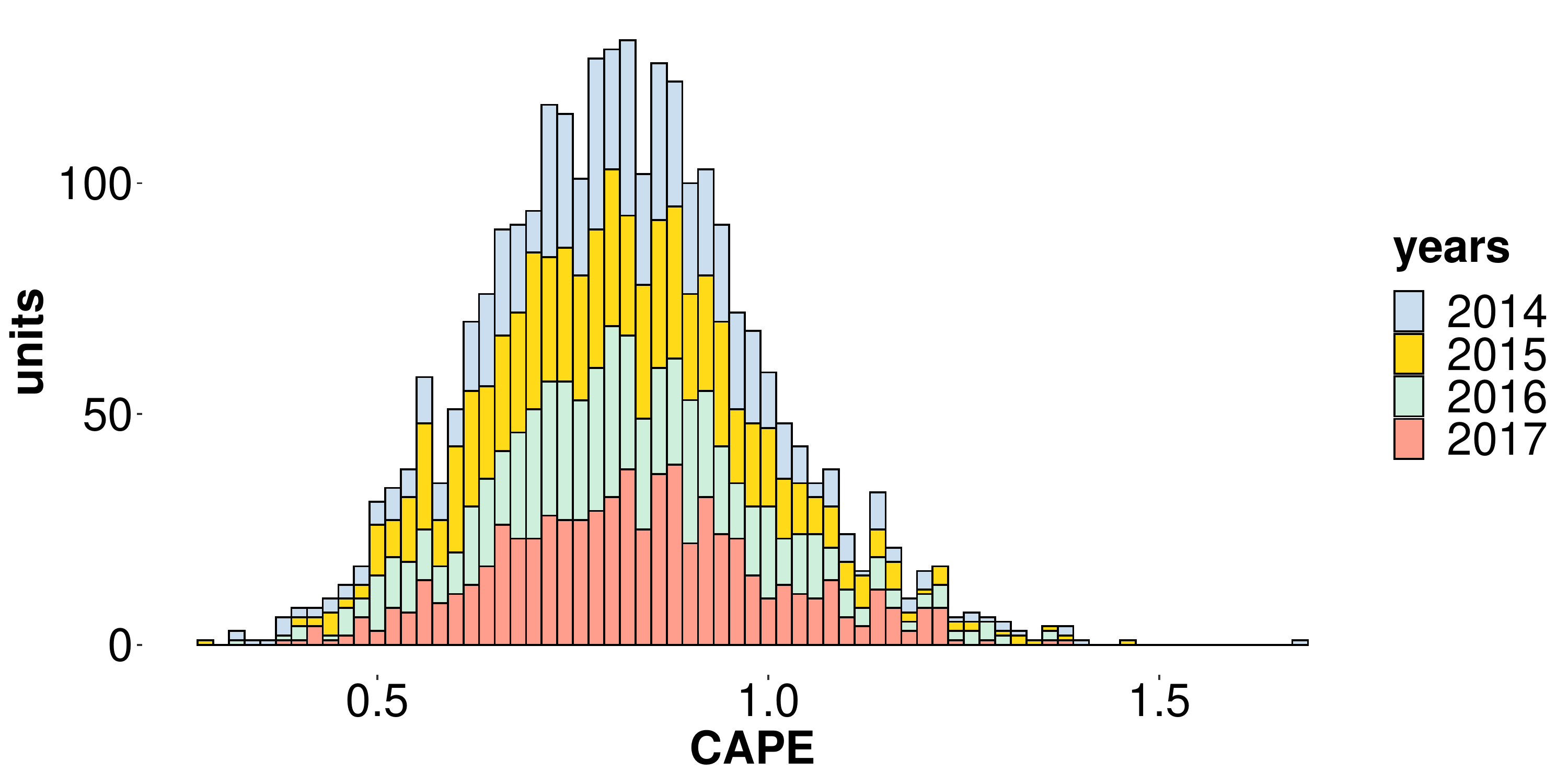}
    \caption{Physician level estimates of payment effects (CAPE) on brand drug prescriptions aggregated by year. CAPE are measured as changes in brand drug claims for a 10\$ increase in payment.}
    \label{fig:histcapeyears}
\end{figure}

\begin{figure}[H]
    \centering
    \includegraphics[scale=0.4]{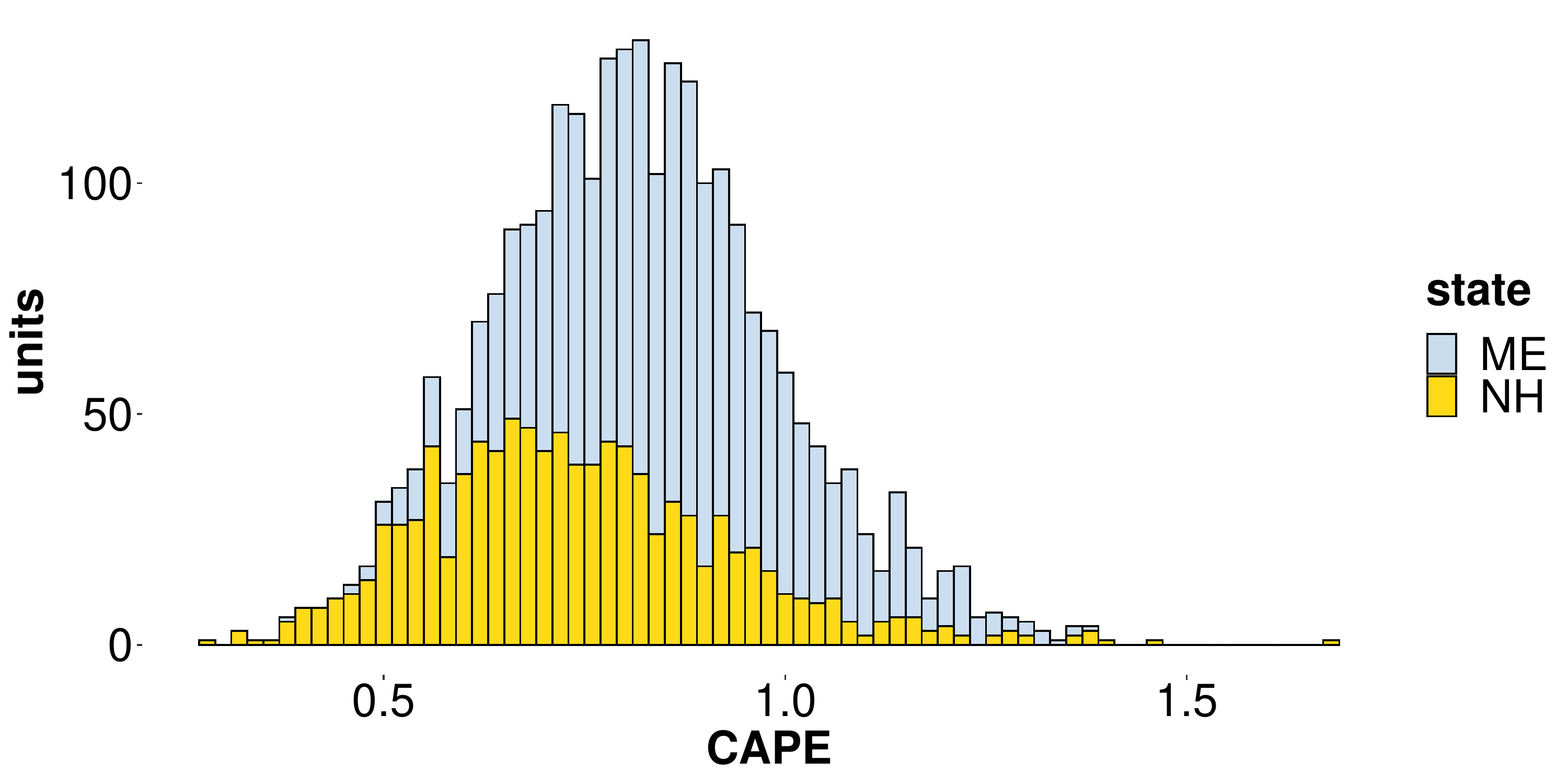}
    \caption{Physician level estimates of payment effects (CAPE) on brand drug prescriptions aggregated by state. CAPE are measured as changes in brand drug claims for a 10\$ increase in payment.}
    \label{fig:histcapestate}
\end{figure}

\end{appendices}

\newpage

\footnotesize
\bibliographystyle{chicago}

\bibliography{main_clean}
\end{document}